\documentclass[prb,floatfix,showpacs,superscriptaddress,longbibliography,twocolumn]{revtex4-1}

\usepackage[USenglish]{babel}

\usepackage{amsmath,amssymb,amsfonts}

\usepackage{bm}

\usepackage{epsfig}
\usepackage{subfigure}

\usepackage[colorlinks=true,linkcolor=blue,citecolor=red]{hyperref}

\newcommand{\equ}[1]
{Eq.~(\ref{#1})}

\newcommand{\figu}[1]
{Fig.~\ref{#1}}

\newcommand{\secu}[1]
{Sec.~\ref{#1}}

\def\=={\equiv}

\def\cG0{{\cal G}_0} 
\def\cG{{\cal G}}

\def\=={\equiv}

\newcommand{\eqn}[1]{(\ref{#1})}
\newcommand{\dagga}{{\phantom{\dagger}}}
\newcommand{\be}{\begin{equation}}
\newcommand{\ee}{\end{equation}}
\newcommand{\dis}{\displaystyle}
\newcommand{\fract}[2]{\frac{\dis #1}{\dis #2}}
\newcommand{\bR}{\mathbf{R}}
\newcommand{\bRp}{\mathbf{R'}}

\begin{document}

\date{\today}

\author{F.~Grandi}
\affiliation{Scuola Internazionale Superiore di Studi Avanzati (SISSA), Via Bonomea 265, I-34136 Trieste, Italy}

\author{A.~Amaricci}
\affiliation{Scuola Internazionale Superiore di Studi Avanzati (SISSA), Via Bonomea 265, I-34136 Trieste, Italy}
\affiliation{CNR-IOM DEMOCRITOS, Istituto Officina dei Materiali, Consiglio Nazionale delle Ricerche, Via Bonomea 265, I-34136 Trieste, Italy}

\author{M.~Capone}
\affiliation{Scuola Internazionale Superiore di Studi Avanzati (SISSA), Via Bonomea 265, I-34136 Trieste, Italy}

\author{M.~Fabrizio}
\affiliation{Scuola Internazionale Superiore di Studi Avanzati (SISSA), Via Bonomea 265, I-34136 Trieste, Italy}

\title{Correlation-driven Lifshitz transition and orbital order in a two-band Hubbard model}

\begin{abstract}
We study by dynamical mean field theory the ground state of a quarter-filled Hubbard model of 
two bands with different bandwidths. At half-filling, this model is known to display 
an orbital selective Mott transition, with the narrower band undergoing Mott 
localisation while the wider one being still itinerant. 
At quarter-filling, the physical behaviour is different and to some extent reversed. 
The interaction generates an effective crystal field splitting, absent in the 
Hamiltonian, that tends to empty the narrower band in favour of the wider one, 
which also become more correlated than the former at odds with the orbital 
selective paradigm. Upon increasing the interaction, the depletion of the 
narrower band can continue till it empties completely and the system undergoes 
a topological Lifshitz transition into a half-filled single-band metal that eventually turns insulating. 
Alternatively, when the two bandwidths are not too different, a first order Mott 
transition intervenes before the Lifshitz's one. The properties of the Mott insulator 
are significantly affected by the interplay between spin and orbital degrees of freedom.  
 
\end{abstract}
\maketitle

\section{Introduction}\label{secI}
Orbital degrees of freedom in correlated materials have witnessed a revived
 interest in recent years mainly motivated by the physics of ruthenates 
 \cite{Anisimov2002-CaSrRuO,PhysRevLett.84.2666}, of iridates and other transition metal 
 compounds with strong spin-orbit coupling \cite{Jackeli-PRL2009,SOC-2016}, 
 and of iron pnictides\cite{Medici2011PRL,Georges2013ACMP,deMedici2014-PRL-Iron}. 
 Realistic lattice Hamiltonians are characterised by tight-binding parameters 
 generically not invariant under orbital $O(3)$ rotations. However, the sensitivity 
 to such orbital symmetry breaking terms depends significantly on the degree of 
 correlations, quantified  by the strengths both of the monopole Slater integral, 
 i.e. the conventional Hubbard $U$, as well as of the higher order multipoles 
 responsible of Hund's rules. For instance, the distinction between different 
 orbitals brought about by the hopping integrals and 
the crystal field can be amplified by strong correlations, 
leading to pronounced orbital differentiation
\cite{Kotliar-orbital-diff-2011,Bascones-PRB2012,deMedici2014-PRL-Iron,Lanata-PRL2017}, 
and eventually to the so-called orbital-selective Mott transitions (OSMT)
\cite{Koga-OSMT-PRL2004,Ferrero-OSMT-PRB2005,Medici2005PRB,PhysRevB.87.205135,
PhysRevB.89.085127,PhysRevB.83.205112,PhysRevB.84.195130,1674-1056-25-3-037103,
PhysRevLett.102.126401,PhysRevB.96.125111,PhysRevB.96.125110,
PhysRevLett.118.177002,PhysRevB.94.235110,PhysRevB.79.115119,
PhysRevLett.102.226402,0953-8984-28-10-105602} where 
the orbitals with the narrowest bandwidth localise while the others are still itinerant.      
In addition, orbital degrees of freedom are expected to play an important role 
in determining which symmetry-broken phase is more likely to accompany the 
Mott transition when correlations grow at integer electron density. This issue 
has been studied quite intensively deep inside the Mott insulator, where one 
can map the Hamiltonian onto a Kugel-Khomskii type
\cite{kugel_khomskii_1973,0038-5670-25-4-R03,PhysRevLett.91.090402,1402-4896-72-5-N02} 
of spin-orbital Heisenberg model\cite{Jackeli-PRL2009}, while it is to a large 
extent unexplored right at the Mott transition. 

In this work we tackle this issue and analyse how orbital degrees of freedom 
affect the zero temperature Mott transition in the simple two-band Hamiltonian where OSMT 
was first observed~\cite{Koga-OSMT-PRL2004}, though at quarter~\cite{1742-6596-150-4-042128} rather than 
at half-filling~\cite{PhysRevB.55.R4855,PhysRevB.66.115107,PhysRevB.81.035112}. 
We will show that, despite its simplicity, this model acquires 
quite a rich phase diagram thanks to the orbital degrees of freedom and 
their interplay with the spin ones.  
The article is organised as follows. In section \ref{secII} we introduce the 
model and anticipate its possible phases by simple weak and strong coupling 
arguments. In sections \ref{secIII} and \ref{secIV} we present the solution of 
the model on a Bethe lattice with infinite coordination number through the 
dynamical mean-field theory. In particular, in section \ref{secIII} we discuss 
the results obtained by preventing magnetic long-range order, which we 
instead allow in section \ref{secIV}. Finally, section \ref{secV} is devoted 
to concluding remarks.

\section{The model}\label{secII}
We consider the Hubbard model of two orbitals with different hopping integrals 
\begin{equation} \label{hamiltonian}
\begin{split}
\mathcal{H} &= - \fract{1}{\sqrt{z}}\sum_{\langle \bR \bRp \rangle,\sigma}\;
\sum_{a=1}^2\;t_{a} \Big( c^{\dagger}_{\bR a \sigma}
  c^\dagga_{\bRp a\sigma} + H.c. \Big)  \\
&\quad + \frac{U}{2} \sum_{\bR} \,n_\bR\,\big(n_\bR-1\big) 
-\mu\sum_{\bR}\,n_{\bR} \;, 
\end{split}
\end{equation}
on a Bethe lattice of coordination number $z$ that we shall eventually send to infinity.  
In \eqn{hamiltonian} the operator $c^\dagga_{\bR a\sigma}$ 
($c^\dagger_{\bR a\sigma}$) annihilates (creates) an electron at site $\bR$ in  
orbital $a= 1,2$ with spin $\sigma =\uparrow,\downarrow$, $n_\bR = \sum_{a \sigma} n_{\bR a\sigma} = \sum_{a\sigma}\,
c^\dagger_{\bR a\sigma}c^\dagga_{\bR a\sigma} $ is the
number operator at site $\bR$,  $\mu$ the chemical potential, and $t_{a}$ a 
nearest neighbour hopping integral, diagonal in the orbital index $a$.  
Hereafter we shall assume $t_1\geq t_2$ and define the hopping anisotropy 
parameter $\alpha = t_2/t_1\in \left[ 0,1 \right]$.

At half-filling, i.e. an average occupation of two electrons per site, 
$\langle n_\bR\rangle=2$,  the Hamiltonian \eqn{hamiltonian} was first 
studied as the simplest toy model to uncover the physics of OSMT
\cite{Koga-OSMT-PRL2004,Ferrero-OSMT-PRB2005,Medici2005PRB}. 
The interaction $U$ makes the narrower band more correlated than the 
wider one, as one would na{\"\i}vely expect, to such an extent that band 
2 may become Mott localised despite band 1 is still itinerant. This 
phenomenon is paradigmatic of many physical situations, the best 
known examples being heavy fermions~\cite{Vojta2010-Heavy-Fermions} 
and ruthenates \cite{Anisimov2002-CaSrRuO}. 

Here we shall instead focus on the quarter-filled density case, 
i.e. $\langle n_\bR\rangle=1$. 
We consider an interaction term (see \eqn{hamiltonian}) which includes the 
monopole Slater integral $U>0$, but not the Coulomb exchange $J$ 
responsible of Hund's rule. This term corresponds to the
density-density part of the Kanamori interaction~\cite{Kanamori1963POTP,Georges2013ACMP}
with no Hund's coupling.

We introduce the local spin and orbital pseudo-spin operators,
$\bm{\sigma}_{\bR}$ and $\bm{\tau}_{\bR}$, respectively, through:
\[
\begin{split}
& \bm{\sigma}_{\bR} =  \sum_{a\sigma\sigma'}
c^{\dagger}_{\bR a\sigma} \;\bm{\sigma}_{\sigma\sigma'}\;
c^\dagga_{\bR a\sigma'}\;, \\ 
& \bm{\tau}_{\bR} = \sum_{\sigma a b}
c^{\dagger}_{\bR a\sigma} \;\bm{\sigma}_{ab}\;
c^\dagga_{\bR b\sigma} \;,
\end{split}
\]
where $\bm{\sigma}=\left(\sigma^x,\sigma^y,\sigma^z\right)$, with
$\sigma^{x,y,z}$ being the Pauli matrices. 
The Hamiltonian \eqn{hamiltonian} is invariant under global spin-$SU(2)$ 
rotations. On the contrary, orbital $SU(2)$ symmetry holds only at $\alpha=1$, 
while for any $\alpha<1$ the symmetry is lowered down to $U(1)$, which 
corresponds to uniform rotations around the orbital pseudo-spin $z$-axis. It 
follows that a finite expectation value of the $z$-component of the uniform 
pseudo-spin operator, which defines the orbital polarisation  
\be
\tau^{z} = \fract{1}{V}\sum_{\bR\sigma}\,\langle\, 
n_{\bR 1\sigma} - n_{\bR 2\sigma} \,\rangle\,,
\ee
$V$ being the number of lattice sites, is allowed by symmetry, 
while a finite expectation value of $\bm{\sigma}_\bR$ and of 
$\tau^{x,y}_\bR$ would break a Hamiltonian symmetry, the 
spin $SU(2)$ and the orbital $U(1)$, respectively. 
We underline that when $\alpha = 1$ the symmetry of the model 
is enlarged to $SU(4)$\cite{PhysRevLett.107.215301}, but in what follows we shall not consider 
such special point.

\subsection{DMFT solution}
We study the model Hamiltonian \eqn{hamiltonian} by means of dynamical 
mean-field theory (DMFT). This is a non-perturbative method that provides 
an exact solution in the limit of infinite lattice-coordination $z\rightarrow\infty$.
\cite{PhysRevLett.62.324,Georges1996RMP} The non-interacting density-of-states corresponding 
to nearest-neighbour hopping $t_a/\sqrt{z}$, $a=1,2$, reads 
\be
\mathcal{D}_{a}(\epsilon) = \fract{2}{ \;\pi D_{a}^2\;} \;
\sqrt{\,D_a^2-\epsilon^2\;}\;,
  \ee 
where $D_{a} = 2 t_{a}$ is half the bandwidth. Hereafter, we shall take 
$D_1=1$ as energy unit, so that $D_2=\alpha \leq 1$. We observe that, since 
the Bethe lattice is bipartite and the Hamiltonian is not frustrated, the most likely 
spatial modulation breaks the symmetry between the two sub-lattices, which we 
shall label as sublattice $\Lambda=A$ and $\Lambda=B$. 
Within DMFT,  the lattice model is mapped onto two distinct
effective impurity problems, one for each sub-lattice. 
Each impurity is coupled to a self-consistent bath, which
is described by a frequency dependent matrix Weiss field 
$\hat{\mathcal{G}}^{-1}_{0\Lambda}(i\omega_n)$, whose matrix elements refer to 
spin and orbital indices.  
Each Weiss field is determined self-consistently
by requiring the impurity problems to reproduce the
local physics of the lattice model, which corresponds to the 
self-consistency equation:
\begin{equation}\label{self-consistency}
\hat{\mathcal{G}}_0^{-1}(i\omega_n) = \hat{G}_\mathrm{loc}^{-1}(i\omega_n)
+ \hat{\Sigma}(i\omega_n)
\end{equation}
where $\hat{G}_\mathrm{loc}$ is the local 
interacting Green's function of the lattice model, and 
$\hat{\Sigma}(i\omega_n)$ the impurity self-energy matrix. In this work we shall 
employ zero-temperature exact diagonalization as impurity solver 
\cite{Caffarel1994a,PhysRevB.86.115136}, with a total number $N_s = 10$ 
of sites. This corresponds to a discretization of the bath of the effective 
Anderson model in $N_{b} = 8$ levels.\footnote{We performed calculations 
with different number of bath sites for selected points in the phase diagram 
in order to check the convergence of our results. We observed that qualitatively 
and quantitatively identical results are obtained already for $N_{b} = 8$. We 
stress that the bath parameters are determined self consistently in order to 
fulfill the DMFT self consistency equations, so they are adaptive. This reduces 
a lot the finite size effects on the solution of the problem, making this approach 
accurate already with relatively few bath levels.}

\subsection{Weak and strong coupling analyses}

\begin{figure}
 \centering \includegraphics[width=0.5\textwidth]{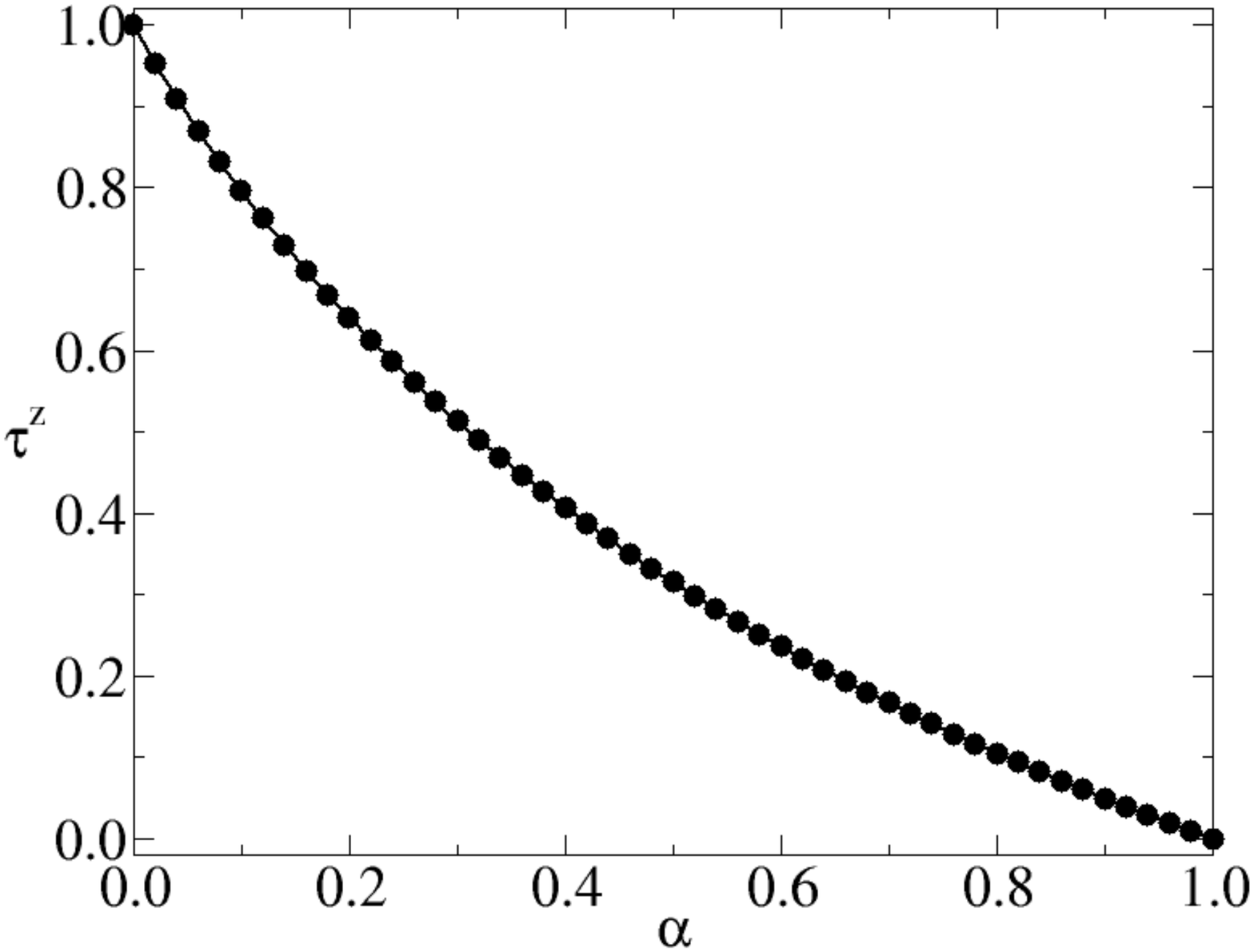}
  \caption{ (Color online) Orbital polarization $\tau^{z}$ as function of $\alpha$ 
  for the non interacting ($U = 0$) case.} 
  \label{fig_tau_0_AFO}
\end{figure}

We can actually anticipate some features of the phase diagram by 
simple arguments in the weak and strong coupling regimes, respectively.
\subsubsection{Weak coupling}\label{weak_coupling_analysis}
When $U=0$, the system describes a quarter-filled two-band metal (2BM) with uniform orbital polarisation 
$\tau^z=0$ at $\alpha=1$ that increases monotonically as $\alpha$ decreases (see \figu{fig_tau_0_AFO}).  
A finite $U\ll \alpha$, small enough to justify the Hartree-Fock approximation, introduces 
an effective crystal field splitting between the two bands 
\begin{equation}
\begin{split}
\mathcal{H} &\to \mathcal{H}_\text{HF} = 
- \fract{1}{\sqrt{z}} \sum_{\langle \bR \bRp \rangle,\sigma}\;
\sum_{a=1}^2\;t_{a} \Big( c^{\dagger}_{\bR a \sigma}
  c^\dagga_{\bRp a\sigma} + H.c. \Big)  \\
&\quad - \sum_{\bR} \,\bigg(\mu_\text{HF}\;n_\bR + 
\Delta^\text{eff}_\bR\,\Big(n_{1\bR}-n_{2\bR}\Big)\bigg)
\;,
\end{split}\label{H-HF}
\end{equation}
where\cite{PhysRevB.87.205108,PhysRevB.91.115102}
\be
\Delta^\mathrm{eff}_\bR=\fract{U}{2}\,\langle\, n_{1\bR}-n_{2\bR}\,\rangle = 
\fract{U}{2}\,\tau^z\,,\,\forall\,\bR\,,\label{crystal-field}
\ee 
which, for any $\alpha<1$, favours the occupation of the band 1 that has 
larger bandwidth. If such mean-field result remained valid even at sizeable 
$U$, we would expect a topological Lifshitz transition from a quarter-filled 2BM into a 
half-filled one-band metal (1BM). We note that, as long as the model remains 
in a quarter-filled 2BM phase, it is stable towards a Stoner-like instability with 
modulated magnetic and/or orbital ordering, which, in the present case, is 
expected to corresponds to a translational symmetry breaking where the 
two-sublattice become inequivalent. On the contrary, the half-filled 1BM 
phase should become immediately unstable towards such symmetry 
breaking \cite{PhysRevB.31.4403}, turning the metal phase into an 
insulating one with magnetic and/or orbital ordering. In particular, since the 
hopping is diagonal in the orbital index, we expect a magnetic order that 
corresponds to a simple N\'eel antiferromagnet, where, because of spin 
$SU(2)$ invariance, symmetry can be broken along any spin direction. 
Conversely, the Hamiltonian for any $\alpha<1$ is only invariant under 
orbital $U(1)$ rotations around the pseudo-spin $z$-axis. Therefore, the 
possible orbital orderings cannot be anticipated as simply as for the spin 
ones, and we must resort to some more sophisticated calculation. 
However, since all transitions are expected to occur at finite $U$, there is 
no guarantee that the above mean-field arguments hold, and thus the need 
of DMFT that is able to provide accurate results for any interaction strength. 

\subsubsection{Strong coupling }\label{strong_coupling_analysis}

\begin{figure}
 \centering \includegraphics[width=0.5\textwidth]{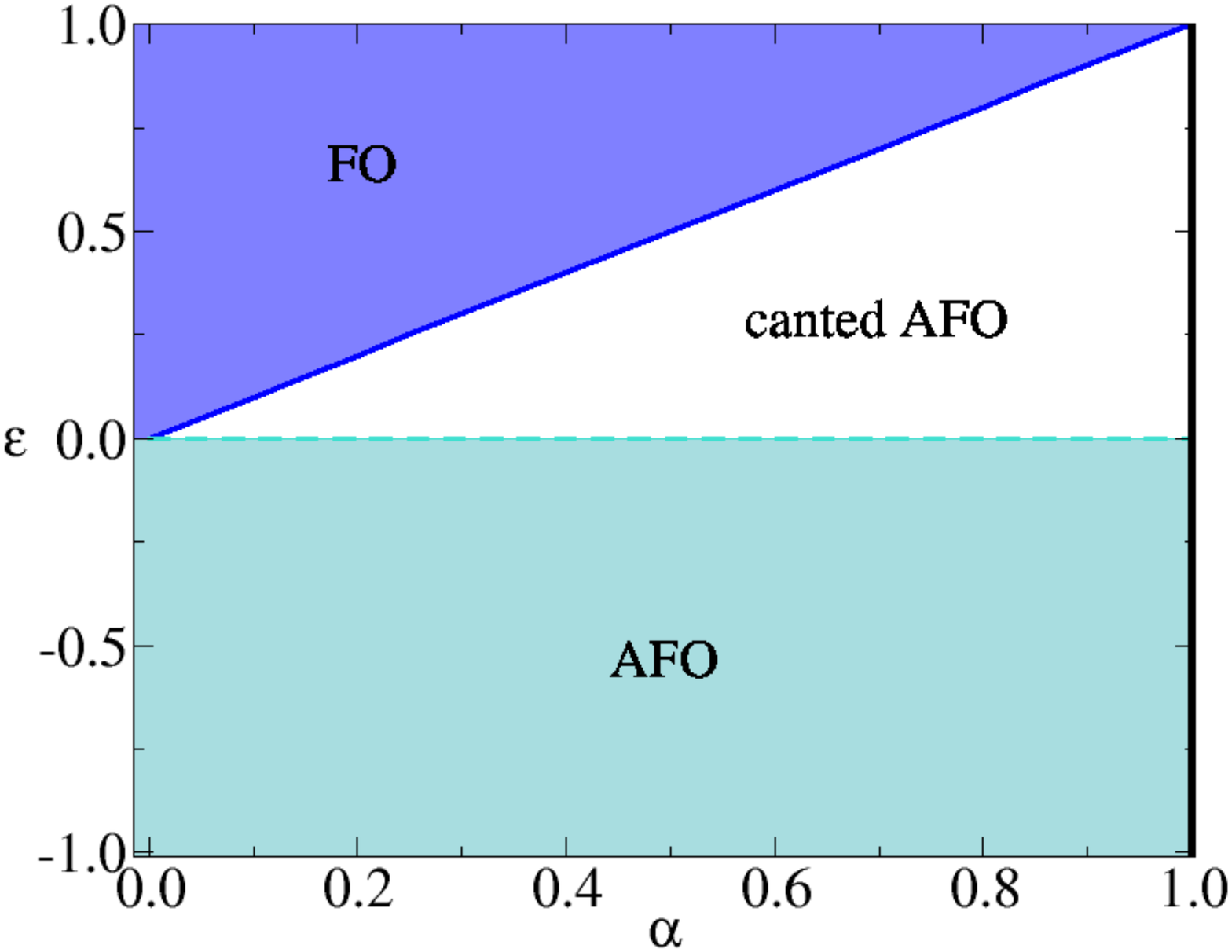}
  \caption{ (Color online) Mean field phase diagram of the strong
    coupling Hamiltonian \equ{strong_coupling_hamiltonian} as a function of $\alpha$ and of the 
  phenomenological parameter $\varepsilon$, defined in \equ{para_cond}. 
  The diagram shows three distinct phases: a ferro- (FO) and an antiferro- (AFO) 
  orbital state along the $z$-direction of the pseudospin and a canted AFO. 
  The AFO phase is connected to the canted AFO through a first order 
  transition (dashed line). The FO phase is separated from the canted 
  AFO by a continuous transition (solid line). 
  When $\varepsilon > 0$ ($\varepsilon < 0$) the system 
  has antiferromagnetic (ferromagnetic) correlations. 
  Along the line $\alpha = 1$ the model is $SU(4)$ invariant, and our simple mean 
  field approximation does not apply any more.} 
  \label{fig_phase_diagr_strong_coupl}
\end{figure}

\begin{figure}
 \centering \includegraphics[width=0.5\textwidth]{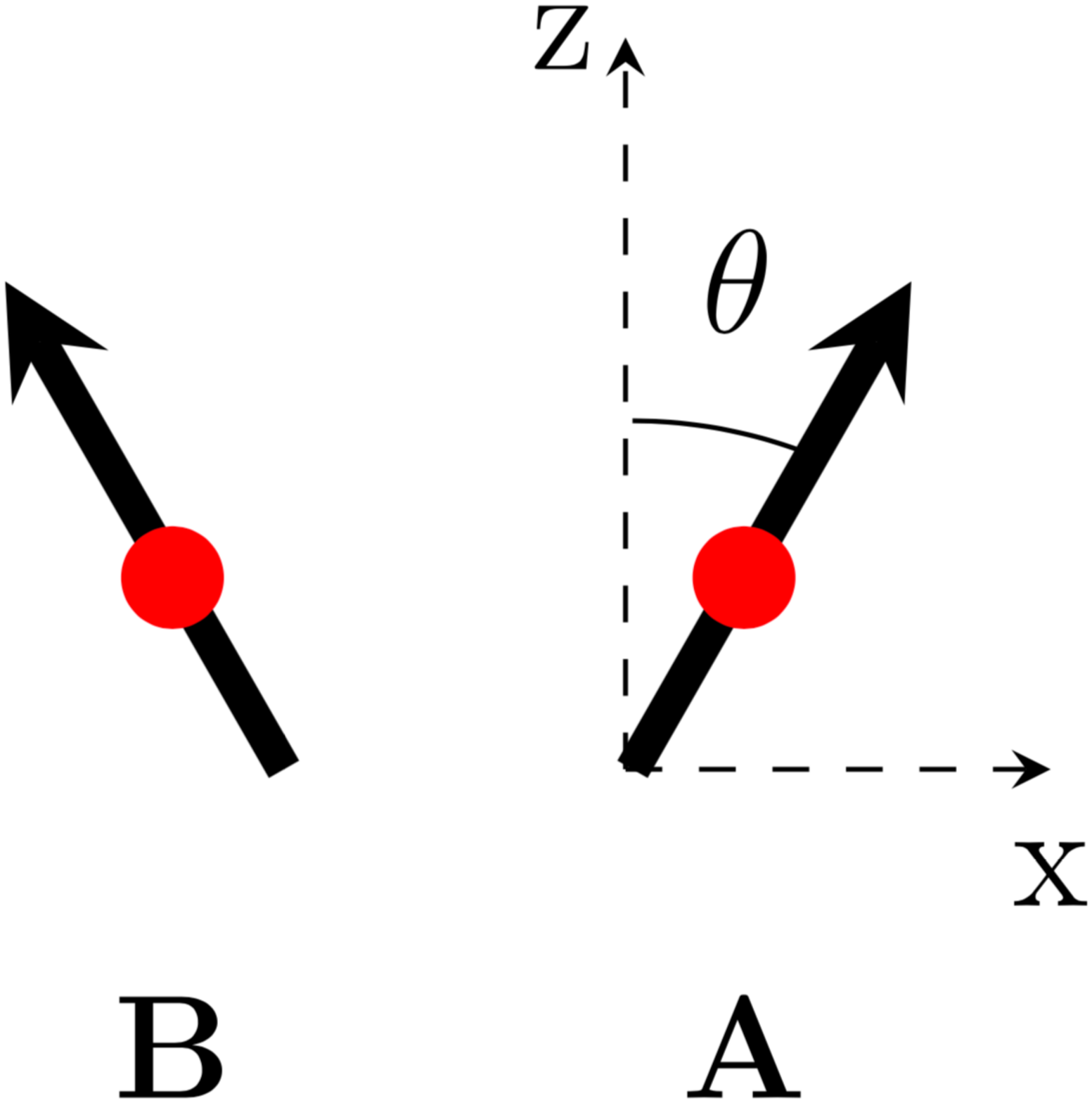}
  \caption{ (Color online) Schematic representation of the canted AFO phase, assuming that 
  the $U(1)$ symmetry is broken along $x$, i.e., $\phi=0$ in \equ{canted AFO}. The
    arrows represent the configuration of the orbital pseudo-spin
    vectors ${\bm \tau}$ at the two sites (red dots) $A$ and $B$ in the unit
    cell. $\theta$ is the angle between the $z$ direction and the pseudospin
    ${\bm \tau}$ on sublattice $A$ (on sublattice $B$ the angle has the value $- \theta$).} 
  \label{fig_canted_afo}
\end{figure}

In order to foresee which orbital ordering is most likely to occur, we can still 
perform some simple analysis. Deep in the Mott insulator, i.e. at strong coupling 
$U\gg 1$, we can map the lattice model \equ{hamiltonian} onto an effective 
Kugel-Khomskii spin-orbital Heisenberg Hamiltonian 
$\mathcal{H}\overset{U\gg 1}{\longrightarrow}\mathcal{H}_\text{KK}$\cite{PhysRev.115.2,kugel_khomskii_1973}, where
\begin{equation}\label{strong_coupling_hamiltonian}
\begin{split}
\mathcal{H}_\text{KK} &= 
\fract{1}{z}\,\sum_{\langle \bR\bRp \rangle}\Bigg\{
\frac{1}{16U} \Big( 1 +
  \bm{\sigma}_\bR \cdot \bm{\sigma}_\bRp \Big)  \bigg[
\big( 1 +\alpha^{2}  \big)   \\
&\quad + \big( 1 - \alpha^{2} \big) \Big( \tau_{\bR}^{z}
  + \tau_{\bRp}^{z} \Big) + \big( 1 + \alpha^{2}  \big)\,
\tau_{\bR}^{z} \,\tau_{\bRp}^{z} \\
&  \quad + 
 2\alpha\, \Big( \tau_{\bR}^{x}\, \tau_{\bRp}^{x} +
  \tau_{\bR}^{y} \,\tau_{\bRp}^{y} \Big)  \bigg] \\ 
& - \frac{1}{8U} \big( 1 - \alpha^{2} \big) \Big(
  \tau_{\bR}^{z} + \tau_{\bRp}^{z} \Big)  -\frac{
  1}{4U} \big( 1 + \alpha^{2}  \big)\,\Bigg\}\;.
\end{split}
\end{equation}
We can solve this hamiltonian at the mean field level factorising the wavefunction into a spin part, 
$\mid \psi_\sigma\rangle$, and an 
orbital pseudo spin one, $\mid \psi_\tau\rangle$. We assume that the expectation value on the spin wavefunction   
\be\label{para_cond}
\langle\psi_\sigma\mid \bm{\sigma}_\bR \cdot \bm{\sigma}_\bRp \mid\psi_\sigma \rangle = - \varepsilon
\in [-1,1]\,.
\ee
Let us briefly comment about the meaning of \equ{para_cond}.  In a generic lattice 
\be\label{mean_value}
\langle \bm{\sigma}_\bR \cdot \bm{\sigma}_\bRp \rangle = 
\langle \bm{\sigma}_\bR \rangle \cdot\langle \bm{\sigma}_\bRp \rangle + \mathcal{O} \left( \fract{1}{z} \right)\,, 
\ee
so that in the limit of infinite coordination, $z\to\infty$, the parameter $\varepsilon$ in \equ{para_cond} 
is finite as long as spin $SU(2)$ symmetry is broken, in which case the mean-field approximation predicts 
an antiferromagnetic spin configuration, $\varepsilon=1$, and a ferro-orbital (FO) one, with 
expectation value $\langle \psi_\tau \mid \tau^z_\bR\mid
\psi_\tau\rangle = 1$, $\forall\, \bR$. On the contrary, if we were to discuss the mean-field phase diagram of the Hamiltonian (\ref{strong_coupling_hamiltonian}) in the paramagnetic sector and in the limit $z\to\infty$, we should, strictly speaking, set $\varepsilon=0$. In this case the mean-field approximation for 
any $0<\alpha<1$ predicts two degenerate pseudo spin configurations, one, which we denote as antiferro-orbital (AFO), characterised by the finite expectation value $\langle \psi_\tau \mid \tau^z_\bR\mid
\psi_\tau\rangle = (-1)^R$, and the other, which we denote as canted antiferro-orbital (canted AFO), 
see \figu{fig_canted_afo}, with non-zero expectation values
\be
\begin{split}
\langle \psi_\tau \mid \big(\cos\phi\,\tau^x_\bR + 
\sin\phi\,\tau^y_\bR\big)\mid\psi_\tau\rangle &= (-1)^R\,\tau^{||}\,,\\
\langle \psi_\tau \mid \tau^z_\bR\mid
\psi_\tau\rangle &= \tau^z\,,
\end{split}\label{canted AFO}
\ee
where $\tau^z=\cos\theta=(1-\alpha)/(1+\alpha)$, $\tau^{||}=\sin\theta$ and $\phi$ is free, signalling breaking of the orbital $U(1)$ symmetry. This result does not agree with DMFT, see below, which suggests that higher order terms in 
$1/U$, not included in \equ{strong_coupling_hamiltonian}, split the above accidental degeneracy. 
As a matter of fact, the actual DMFT phase diagram can be still rationalised through the mean-field treatment of the simple Hamiltonian  (\ref{strong_coupling_hamiltonian}), proviso a finite $\varepsilon$ is assumed even in the paramagnetic sector and despite $z\to\infty$. \\
For the above reason, we shall hereafter take $\varepsilon$ as a free parameter, in terms of which the phase diagram as function of $\alpha$ is that shown in \figu{fig_phase_diagr_strong_coupl}.
Whenever $\varepsilon < 0$ (ferromagnetic correlations) and $\alpha < 1$ 
the system is in an AFO state. 
When instead $\varepsilon > 0$, as physically expected, we find either a FO state for $\alpha <
\varepsilon$ or a canted AFO one otherwise. The transition between the two phases is continuous within mean-field. Finally, for $\varepsilon = 0$, as we mentioned,  the canted  AFO and the AFO are accidentally degenerate. The transition between them is first order. 

\section{Paramagnetic DMFT results} \label{secIII}
We now turn to exact DMFT and start by analysing the model \eqn{hamiltonian} searching for paramagnetic solutions. However, since the Hamiltonian is not orbital
pseudo-spin invariant, we cannot avoid orbital ordering.

\begin{figure}
 \centering \includegraphics[width=0.5\textwidth]{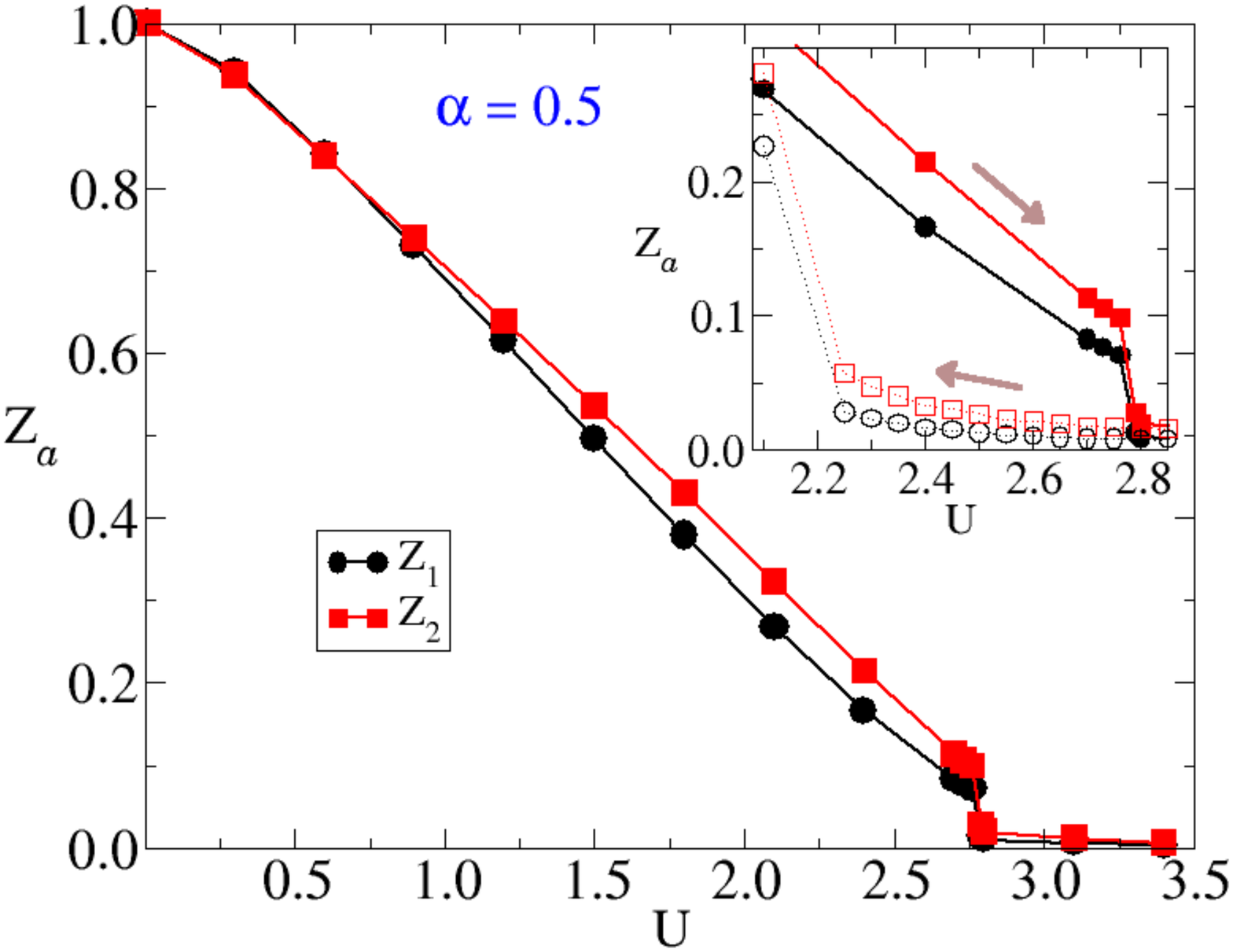}
  \caption{ (Color online) The quasiparticle residues 
    $Z_a$ as function of $U$, for
    $\alpha = 0.5$.  Both $Z_1$ and $Z_2$ vanish at
    $U=U_{c2}\simeq2.80$ signalling transition to the Mott insulator. 
    Inset: Hysteretic behavior of $Z_a$  near the critical point. 
    Filled (open) symbols are obtained continuing the solution from
    small (large) values of $U$.}
  \label{fig_renorm_05_AFO}
\end{figure}

We first consider an intermediate value of the bandwidth ratio $\alpha =
0.5$ and we show how the weakly interacting 2BM is driven
to a Mott insulating state by increasing the interaction strength
$U$. Such phase-transition is revealed  by the 
evolution of the quasiparticle residue
\be
Z_a = \Bigg(
1- \fract{\partial \text{Re}\Sigma_{aa}(\omega)}{\partial\omega}\Bigg)^{-1} _{|\omega = 0}\;,
\ee 
which quantifies the degree of Mott's localization of  
quasi-particles, being $Z_a\to 1$ in the non-interacting limit and $Z_a\to 0$
at the Mott transition.

The results for $Z_a$ are reported in \figu{fig_renorm_05_AFO}. 
In the weakly interacting regime the effects of the interaction are
nearly identical on the two bands, i.e. $Z_1\simeq Z_2$. 
However, upon increasing $U$, the two quantities start
differentiating, with the wider band becoming more correlated than 
the narrower one, i.e. $Z_1<Z_2$\cite{PhysRevA.85.013606}, at odds with the paradigm of the orbital selective Mott 
transition\cite{Koga-OSMT-PRL2004}. 
At a critical value of $U$, the electrons on both
bands localize, as signalled by the simultaneous vanishing of
$Z_1$ and $Z_2$. We find that the metal-insulator Mott transition is first 
order.  In the inset of \figu{fig_renorm_05_AFO} we show that $Z_a$ at the transition suddenly jump 
to zero, and we also observe a clear hysteresis loop.  
The coexistence region extends between $U_{c1}\simeq2.20$ and $U_{c2}\simeq 2.80$.

\begin{figure}
 \centering \includegraphics[width=0.5\textwidth]{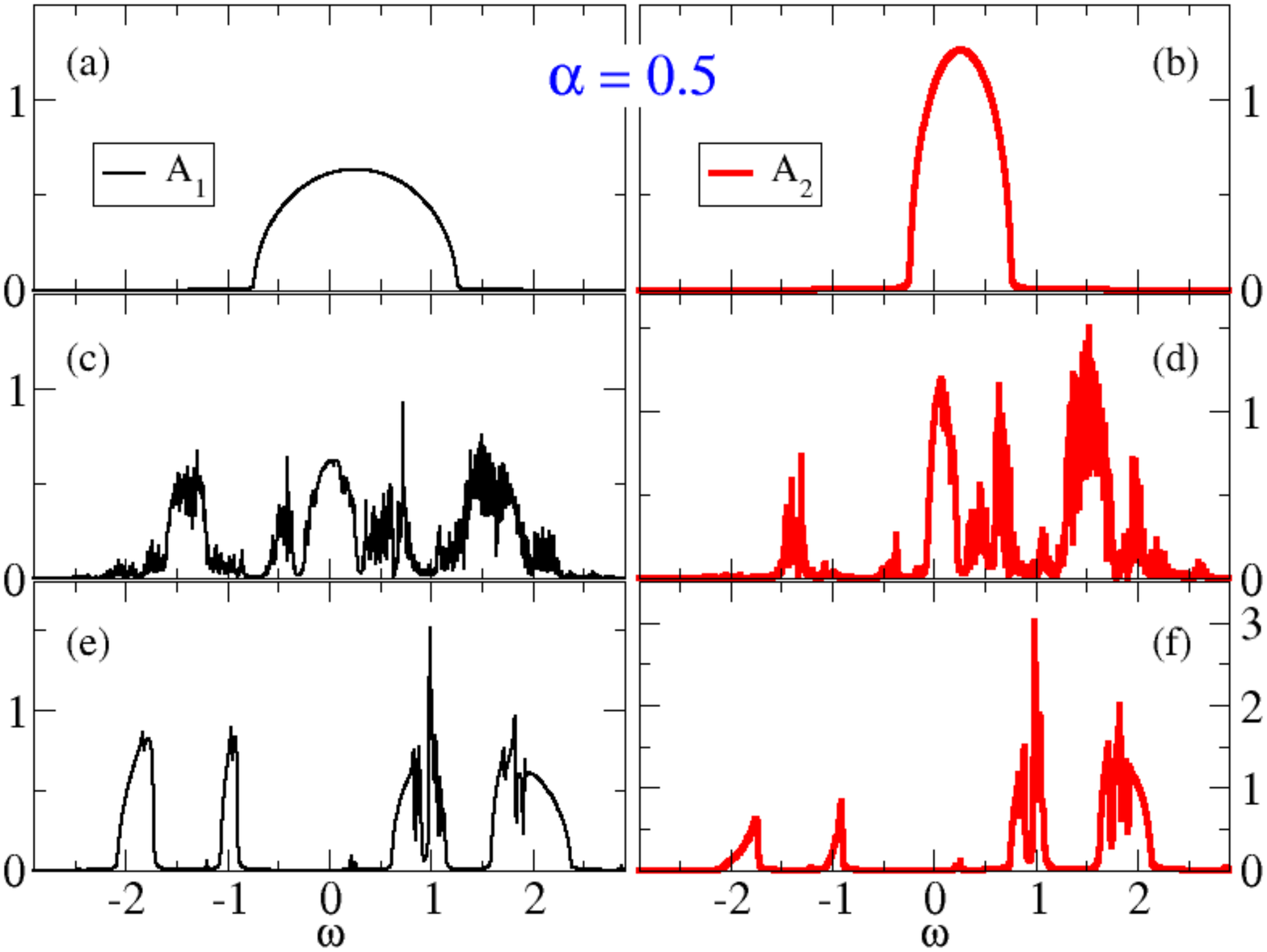}
  \caption{ (Color online) The spectral
    functions $A_{a}(\omega)$ for $\alpha = 0.5$ and sublattice 
    $\Lambda=A$. Data for $a=1$ ($a=2$) are reported on the left
    (right) column. The results are for increasing values of $U$: 
    $U=0.0$ (panels (a), (b)), 
    $U=2.1 < U_{c1}$ (panels (c), (d)), 
    $U=3.1 > U_{c2}$ (panels (e), (f)).} 
  \label{fig_dos_05_AFO}
\end{figure}
A direct insight into the solution is obtained by the 
evolution of the spectral functions $A_{a}\left( \omega \right) = - 
\frac{1}{\pi} \mathrm{Im}{G^{aa}_\mathrm{loc} 
\left( \omega \right)}$ with $a=1,2$,  shown in \figu{fig_dos_05_AFO}. 
At $U=0$ the spectral functions have the
typical semi-elliptical shape of the Bethe lattice.   
Upon increasing the interaction, see \figu{fig_dos_05_AFO}(c)-(d), we observe at high-energy the
gradual formation of the Hubbard sidebands, coexisting with the low-energy quasiparticle peaks. 
For $U>U_{c2}$ the system undergoes a transition into a Mott
insulator. The corresponding spectral functions show a large
gap around the Fermi level ($\omega=0$)  and the two
Hubbard sidebands centred at about $\omega=\pm U/2$. 
\begin{figure}
 \centering \includegraphics[width=0.5\textwidth]{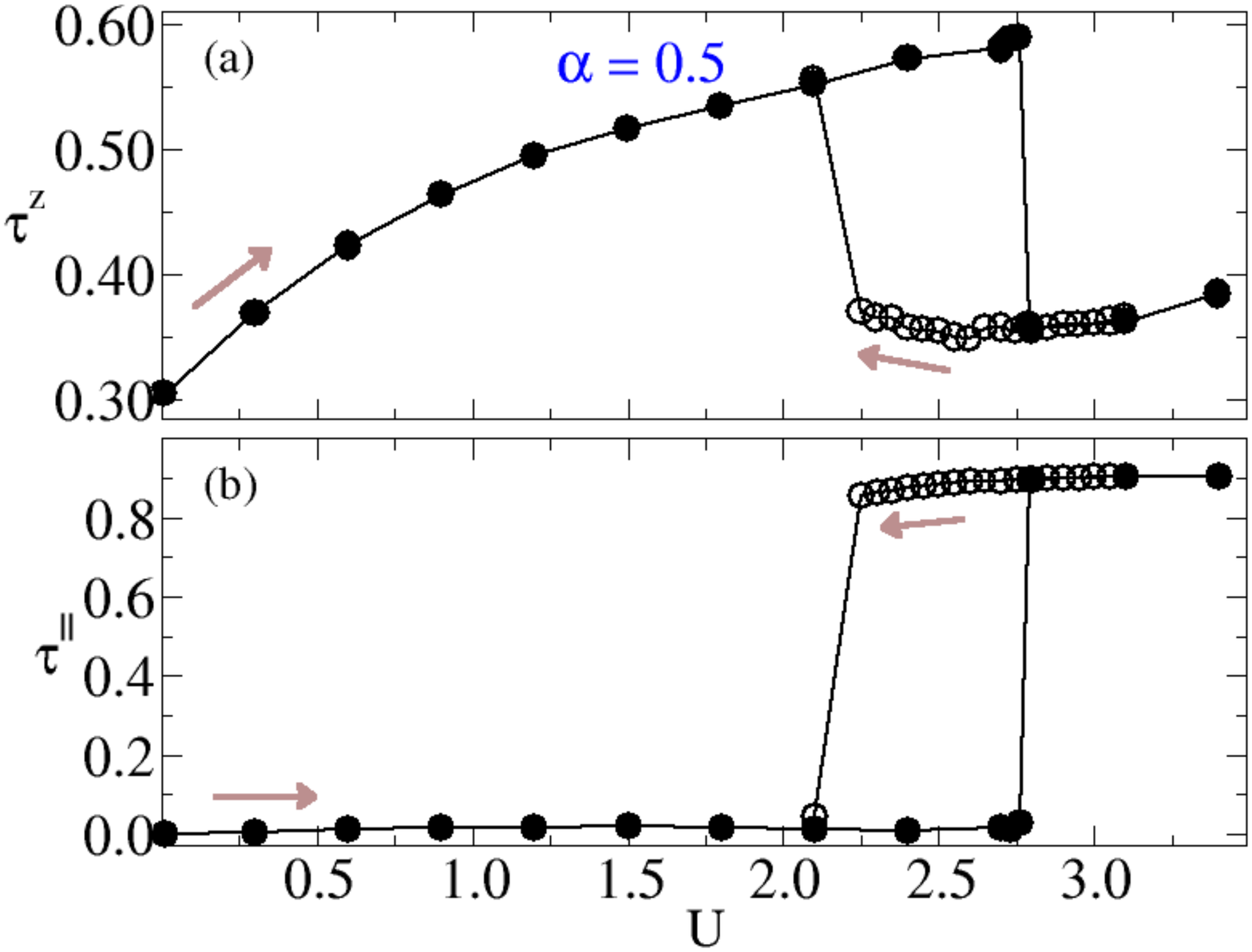}
  \caption{ (Color online) 
    Orbital polarization $\tau^{z}$ (a) and staggered
    in-plane component of the pseudospin $\tau^{||}$ (b) as
    function of the interaction strength $U$. 
    Data are for $\alpha = 0.5$. 
    The arrow indicate the direction in the hysteresis cycle.} 
  \label{fig_tau_05_AFO}
\end{figure}
We note that in the Mott insulator the band 2 has still weight below the Fermi level, namely, 
unlike the mean-field expectation, we do not find a transition into a one-band 
model with maximum orbital polarisation. In \figu{fig_tau_05_AFO} we show the values of 
the uniform orbital polarization, $\tau^z$,  and staggered one, $\tau^{||}$,  as function of 
$U$ across the Mott transition. We always find a finite uniform polarisation, but also an 
antiferro-orbital polarisation in the $xy$-plane, which we have denoted as canted
\textit{AFO} state. This result suggests that the observed degeneracy between the $AFO$ 
along the $z$ direction and the canted $AFO$ mentioned in \secu{strong_coupling_analysis} 
is removed in favor of the canted $AFO$ state.

In the non-interacting limit, $\tau^{||}=0$ while the uniform orbital polarization along $z$ 
is finite, due to the different bandwidths of the two orbitals. In agreement with mean-field, 
upon increasing $U$ the wide band population grows at expenses of the narrow one, 
thus leading to an increase of $\tau^z$ while $\tau^{||}$ remains zero. However this 
tendency does not proceed till a 2BM-to-1BM transition, i.e. till $\tau^z\to1$; before that 
happens a first-order Mott transition takes place. At the transition, we find a sudden 
increase of $\tau^{||}$ to an almost saturated value $\tau^{||} \approx 0.9$, and,  
consequently, $\tau^z$ suddenly drops to a very small value, only slightly larger than 
the non-interacting one. 

\begin{figure}
 \centering \includegraphics[width=0.5\textwidth]{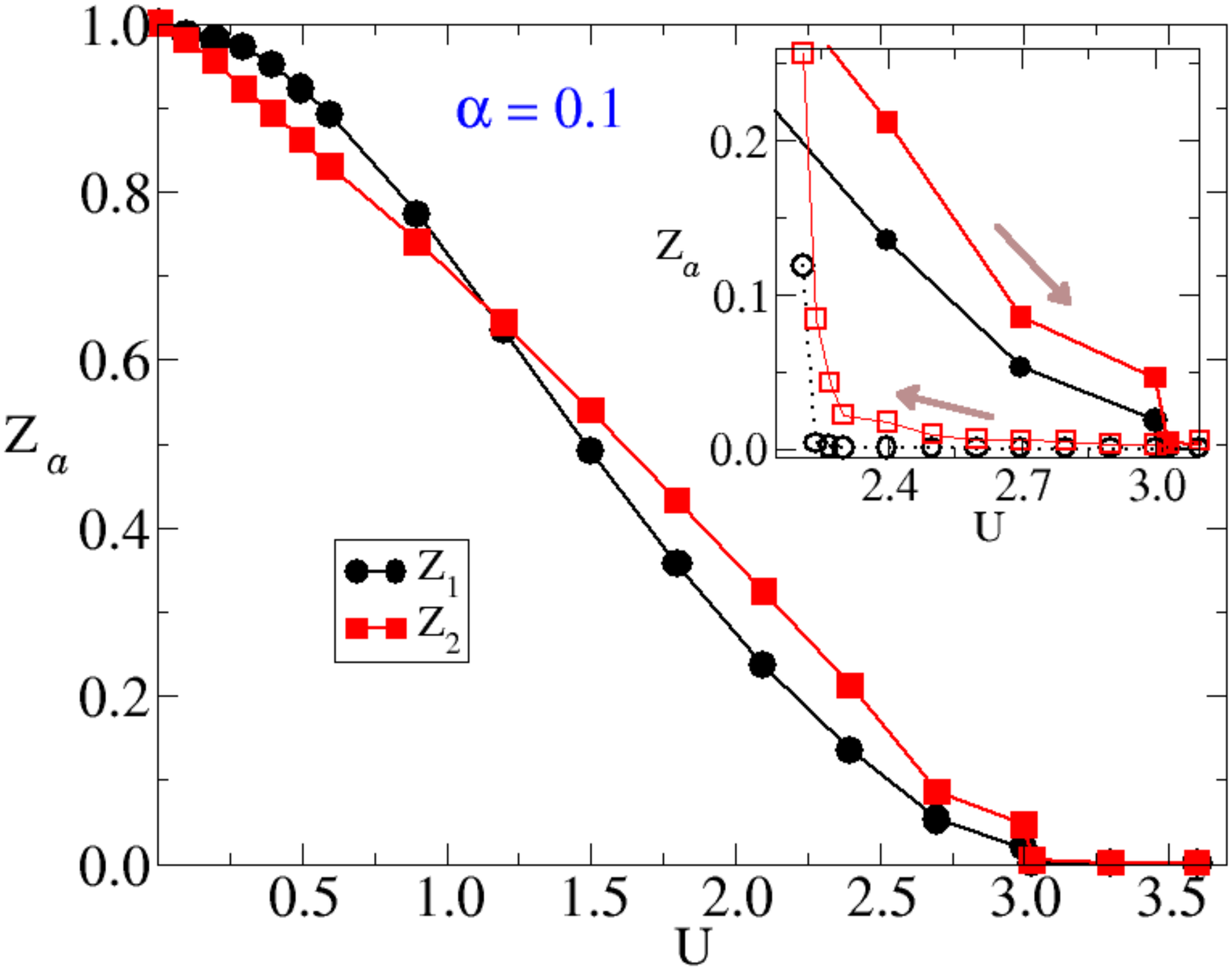}
  \caption{(Color online) Quasiparticle residues $Z_a$ as function of $U$ and for $\alpha=0.1$. 
    Inset: the same quantities near the first order transition. 
    The arrows indicate the hysteresis cycle.} 
  \label{fig_renorm_01_AFO} 
\end{figure}

We now consider a smaller value of the bandwidth ratio, $\alpha = 0.1$. 
The large mismatch between the two bandwidth greatly enhances the
occupation imbalance among the two orbitals, already in the
uncorrelated regime. 
We start by the behaviour of the quasiparticle residues 
$Z_a$, shown in
\figu{fig_renorm_01_AFO}. 
Differently from the previous $\alpha=0.5$ case, the two bands
have distinct $Z_a$ already at relatively small values of $U$, now with 
the narrower band more correlated than the wider one. This behaviour is reversed at
$U\simeq1.2$, at which the wider more populated band 1  
becomes also the most correlated one. 
Further increasing the correlation strength eventually drives the system into
a Mott insulating state, as before through a first-order transition at which both quasiparticle residues drop to zero.

\begin{figure}
 \centering \includegraphics[width=0.5\textwidth]{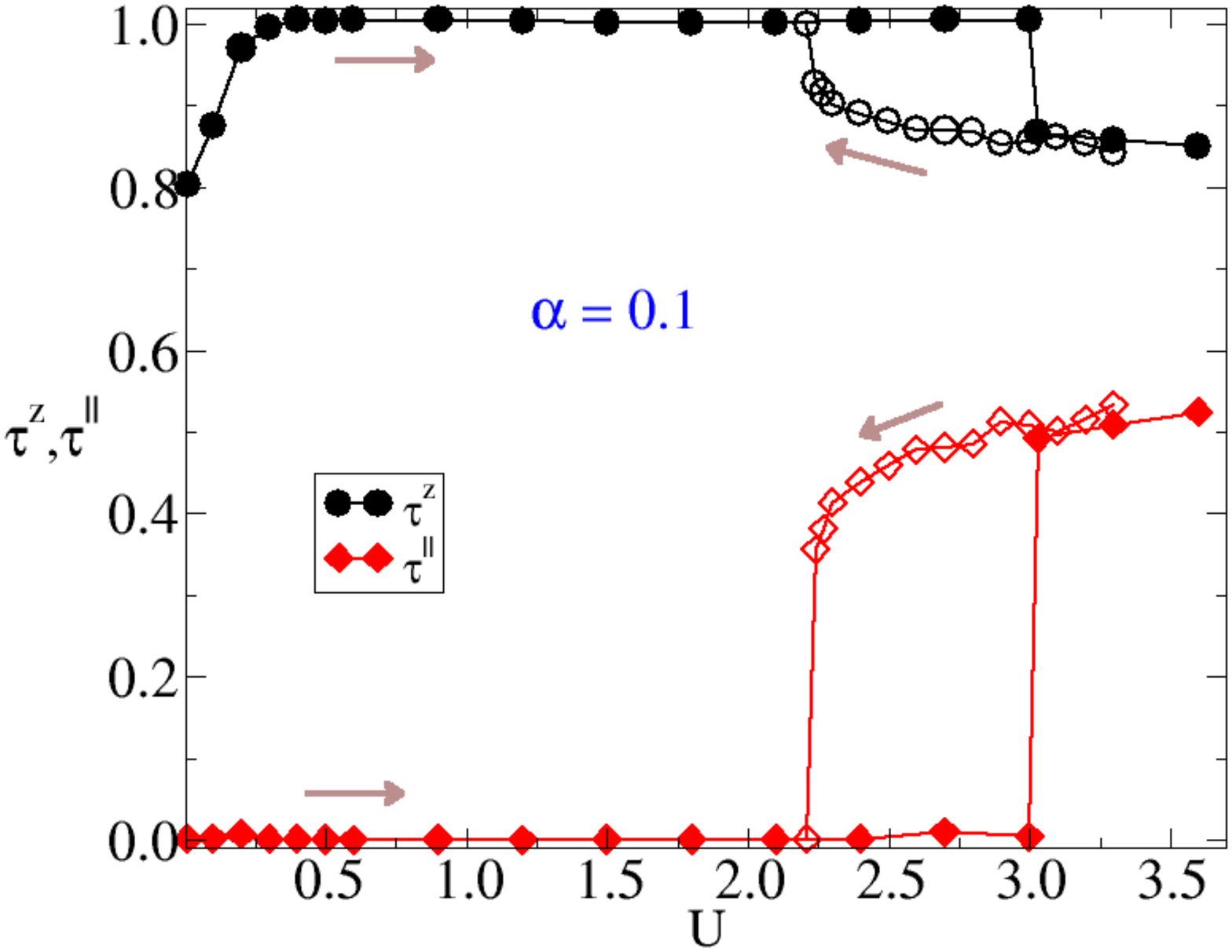}
  \caption{(Color online) Uniform orbital
    polarization, $\tau^{z}$, and staggered
    one, $\tau^{||}$, as a function of
    $U$. Data are for $\alpha = 0.1$. 
    The arrows indicate the hysteresis cycle near the
    Mott transition. } 
  \label{fig_tz_tpar_01_AFO} 
\end{figure}

It is useful to compare the behaviour of $Z_a$ with the evolution of the
orbital polarisations $\tau^z$ and $\tau^{||}$, shown in
\figu{fig_tz_tpar_01_AFO}.  For very small $U$ the system is
characterised by a large value of uniform polarisation, $\tau^z$,  
and vanishing staggered one, $\tau^{||}$. By slightly increasing the interaction strength,  the orbital
polarisation rapidly saturates to $\tau^z=1$. Concomitantly, the narrower band empties while the wider one 
reaches half-filling. Therefore
correlation drives in this case a continuous topological Lifshitz transition from a 2BM to a 1BM, as predicted by the Hartree-Fock approximation. 
Interestingly, the narrower band keeps a high degree of correlations, as demonstrated
by the decreasing behaviour of $Z_2$, 
see \figu{fig_renorm_01_AFO}. In other words, although essentially
empty, the band 2 is not completely decoupled from band 1. 

\begin{figure}
 \centering \includegraphics[width=0.5\textwidth]{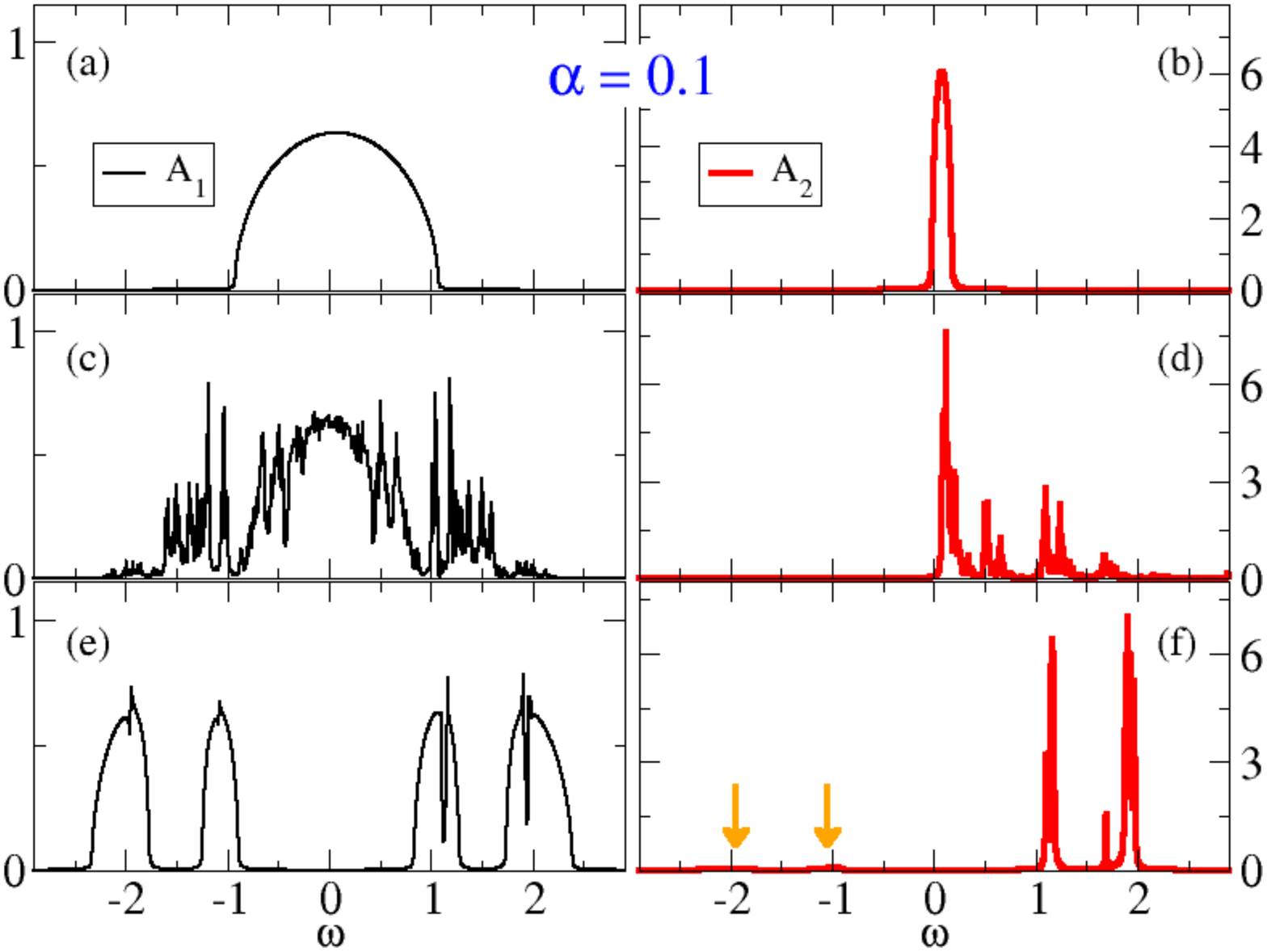}
  \caption{(Color online) Spectral functions
    for $\alpha = 0.1$ and fixed spin on sub-lattice $A$. Data are for
    increasing values of $U$: $U = 0.0$ ((a), (b)), 
    $U = 1.2$ ((c), (d))
    $U = 3.3$ ((e), (f)).
    Note the different scales in the $y$-axis. 
    Arrows in panel (f) indicate tiny spectral weight below the Fermi level 
    for narrow band.} 
  \label{fig_dos_01_AFO} 
\end{figure}

More insights can be gained by the behaviour of the spectral functions, shown 
in \figu{fig_dos_01_AFO}.  The large orbital occupation imbalance
is already visible in the non-interacting limit, with the wider band
being nearly centred around the Fermi level and, 
correspondingly, the narrower one nearly empty. 
Upon increasing the interaction $U$, the narrower band 2 gets shifted entirely
above the Fermi level, yet it still shows spectral weight at high
energy resulting from correlation effects. Simultaneously, the wider band recovers a
particle-hole symmetric shape characterised
by a three-peaks structure, with a
renormalised central feature flanked by the two precursors of the
Hubbard sidebands. 
For $U>U_{c2}$ a spectral gap opens in the the half-filled wider band signalling the onset of 
a Mott insulating state. 
Notably, also the previously empty narrow band shows the formation of a Mott gap
which separates a large spectral feature above the Fermi level from a tiny
spectral weight below it, see the arrows in \figu{fig_dos_01_AFO}(f). 
The systems is thus characterized by $Z_1=Z_2=0$ when it enters into the Mott state, see \figu{fig_renorm_01_AFO}. As for the larger values of 
$\alpha$, the resulting insulating state has a finite in-plane staggered 
polarization, $\tau^{||}$, and a reduced value of the uniform one, $\tau^z$, see \figu{fig_tz_tpar_01_AFO}. 

\begin{figure}
 \centering \includegraphics[width=0.5\textwidth]{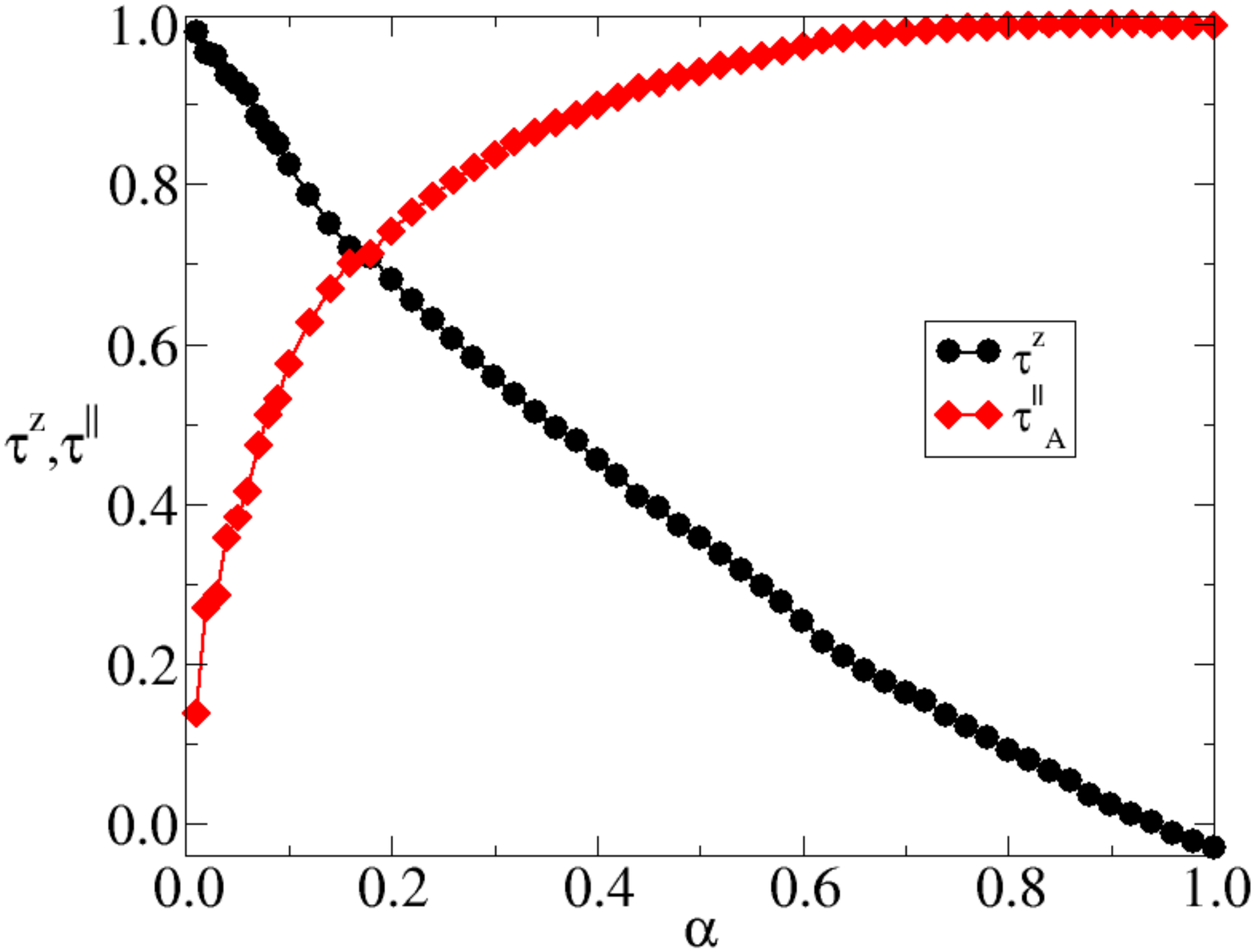}
  \caption{ (Color online) Uniform orbital
    polarization $\tau^{z}$ and staggered
    in-plane component of the pseudospin $\tau^{||}$ as
    function of $\alpha$. Data are for $U = 5.0$.} 
  \label{fig_tau_5_AFO}
\end{figure}

In order to ascertain the strong-coupling picture of section 
\ref{strong_coupling_analysis}, we study the evolution of the orbital 
order in the Mott insulator at large $U$. 
In \figu{fig_tau_5_AFO} we report the behaviour of both uniform, $\tau^{z}$,
and staggered, $\tau^{||}$, polarisations as function of $\alpha$ for  $U = 5$. 
When $\alpha \to 0$, $\tau^{z} \to
1$ and $\tau^{||}\to 0$, while the opposite occurs for 
$\alpha\to 1$. The evolution between these two limits is continuous, namely the critical  
$\alpha_c=0$. We note that those results do not change by decreasing or increasing 
the interaction strength, provided the system remains within the insulating regime. This 
result further confirms the larger stability of the canted $AFO$ with respect to the AFO 
along the $z$ direction in the paramagnetic domain.

\begin{figure}
 \centering \includegraphics[width=0.5\textwidth]{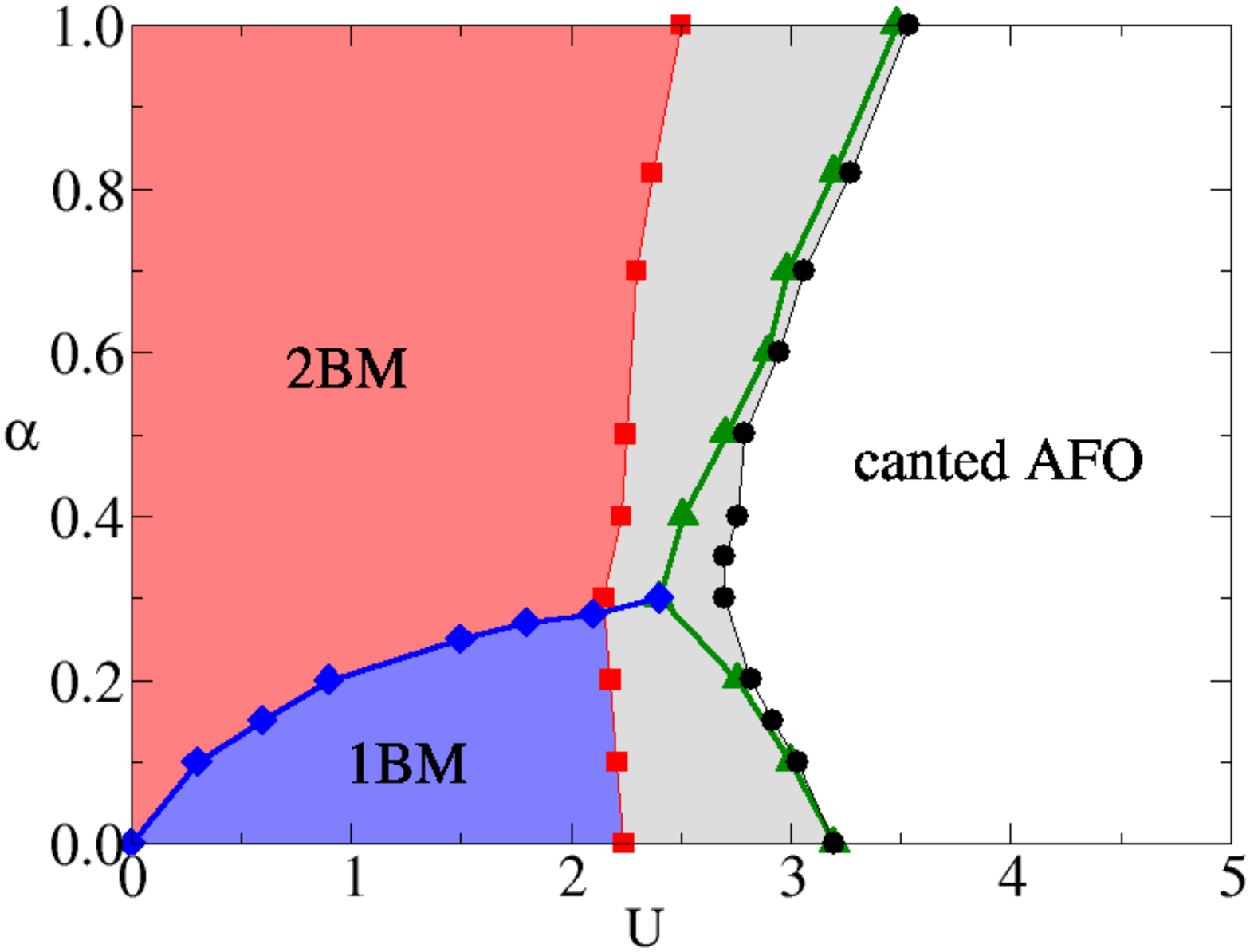}
  \caption{\label{fig_AFO} (Color online) The non-magnetic phase
    diagram of the model in the $U$--$\alpha$ plane. 
    Three different phases are present: 
    a two-bands metal (2BM) at small $U$ and large enough
    $\alpha$; a one-band metal (1BM) for small $\alpha$ and small $U$; 
    and a Mott insulator with canted AFO order. 
    The 2BM phase is connected to the 1BM through
    a continuous topological Lifshitz transition (diamonds).  The transition to the canted AFO ordered 
    Mott insulator is of first-order. 
    The spinodal lines  (filled circles and
    squares) delimitate the coexistence region. 
    The first-order critical line (filled triangles) is computed from
    the energy crossing of the two solutions. A tricritical point is 
    present at the merging of the transition line.} 
\end{figure}

We summarise all previous results in the
$U$-$\alpha$ phase-diagram of \figu{fig_AFO}. 
We find three distinct phases: a metallic state at small
$U$ and large enough $\alpha$ in which  
both bands are occupied (2BM); a metallic phase at small $U$ and
$\alpha$ with a half-filled wider band and an empty narrower one (1BM);
a canted AFO ordered Mott insulator at large
enough interaction. 
The two metallic phases are connected through
a continuous Lifshitz transition\cite{lifshitz_1960} associated to the correlation induced
emptying of the narrow band. 
For a generic value of $\alpha$, increasing the interaction $U$ drives the system into a Mott state
through a first-order transition. This transition is associated with a large
coexistence region (grey shaded area) for
$U_{c1}<U<U_{c2}$\cite{0034-4885-50-7-001}. 
The merging of the Mott and the Lifshitz transition lines is
a tricritical point \cite{PhysRevLett.24.715}.  
Interestingly, the insulator and the 1BM spinodal lines show a residual dependence on $\alpha$. 
This reveals the strong entanglement between the two bands. Thus, although in the 1BM 
phase the wider band is half-filled and particle-hole symmetric, its
description can not be simply reduced to that of a single-band Hubbard model.  
This description is recovered only in the limit $\alpha \to 0$, where
just the broader band is filled for each value of the interaction
strength. We emphasise that the quarter filling condition $\langle n_\bR\rangle=1$
differentiates this model from the Falicov-Kimball
one\cite{PhysRevLett.22.997}. We find that the 1BM to Mott insulator transition at $\alpha=0$ 
takes place continuously at $U_c=U_{c2}$, as
in the DMFT description of the Mott transition in the single-band
Hubbard model\cite{Georges1996RMP}. However, for any non-zero $\alpha$ a finite staggered 
in-plane polarisation appears, and thus both bands are partially occupied.   


\section{Anti-ferromagnetic DMFT results}\label{secIV}

In the previous section we artificially prevented the DMFT solution to spontaneously 
break spin-$SU(2)$ symmetry and order magnetically, specifically into a simple 
N\'eel antiferromagnetic configuration since the lattice is bipartite and the Hamiltonian 
not frustrated. Here we shall instead leave the system free to order also magnetically, 
and study the interplay between spin and orbital orderings. Because of spin $SU(2)$ 
symmetry, all symmetry breaking directions are equivalent, and thus we choose for 
convenience the $z$-axis and define the staggered magnetisation of orbital $a = 1,2$ as 
\[
m_a = \fract{1}{V}\sum_{\bR\in A}\,
\langle\,n_{\bR a\uparrow}-n_{\bR a\downarrow}\,\rangle
-\fract{1}{V}\sum_{\bR\in B}\,
\langle\,n_{\bR a\uparrow}-n_{\bR a\downarrow}\,\rangle\,,
\]
and the full staggered magnetisation as $m=m_1+m_2$.

\begin{figure}
 \centering \includegraphics[width=0.5\textwidth]{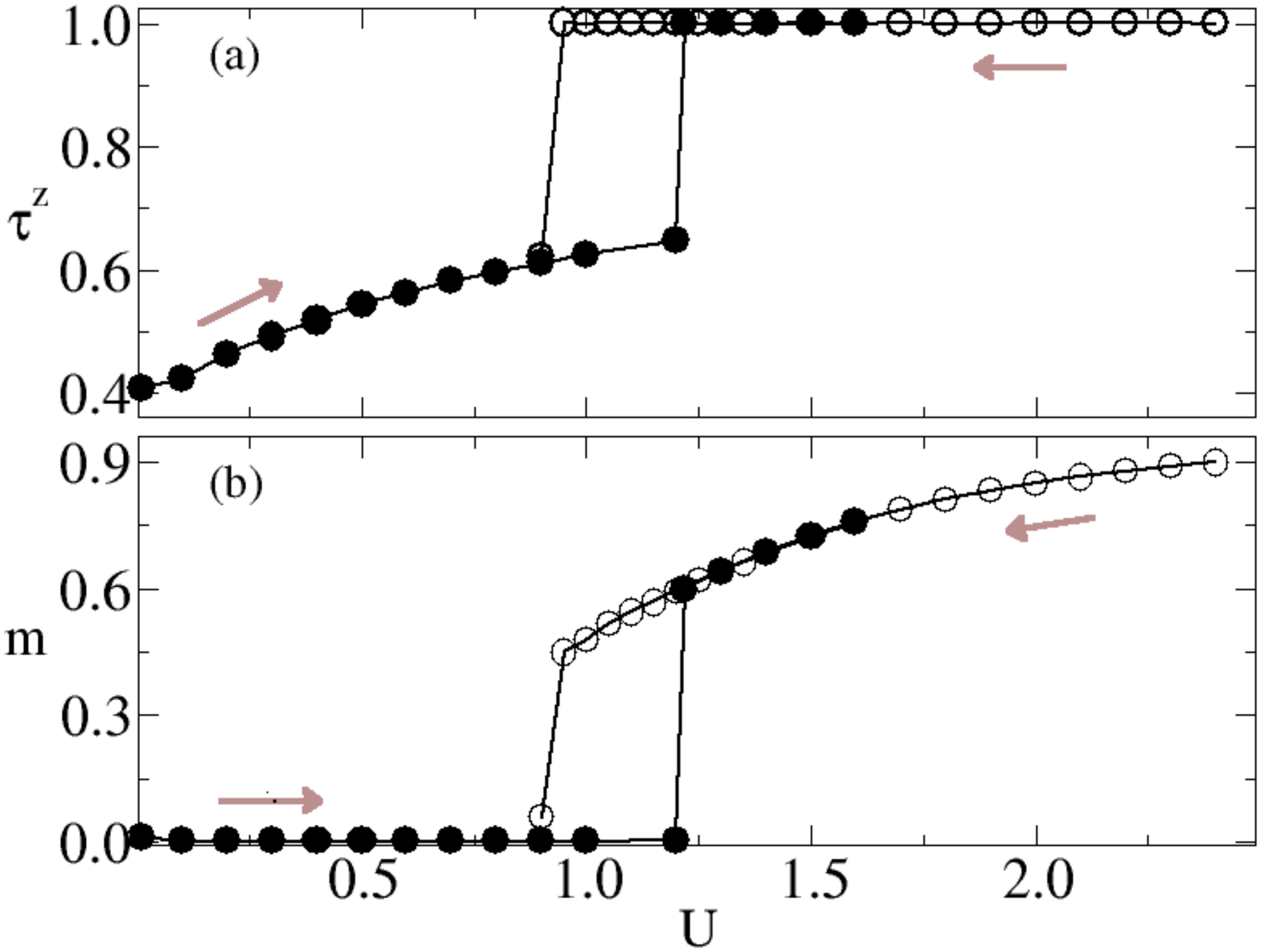}
  \caption{(Color online) Uniform orbital polarization $\tau^{z}$ (a) and
    staggered spin magnetization $m$ (b) as functions of the
    interaction $U$. Data are for $\alpha = 0.4$. 
    The system undergoes a first-order transition from
    the 2BM to an antiferromagnetic (AFM) state, with finite $m$. The 
    orbital polarization saturates to $\tau^z=1$ corresponding to a
    ferro-orbital (FO) ordering of the AFM state.
    The arrows indicate the directions of the solutions in the
    coexistence region $U^{AFM}_{c1}=0.9<U<1.2=U^{AFM}_{c2}$.}
  \label{fig_tau_magn_04_AFM} 
\end{figure}

We start taking $\alpha = 0.4$. In \figu{fig_tau_magn_04_AFM} we
show the evolution of the uniform orbital polarization $\tau^z$ and 
staggered magnetization $m$ as function of $U$.  
By increasing the interaction from $U=0$, $\tau^z$ slowly increases, but 
the system remains a paramagnetic 2BM, thus $m=0$. 
For $U=U^{AFM}_{c2}\simeq 1.2$ we find a first-order
transition to an antiferromagnetic (AFM) ordered state, signalled by the 
sudden increase of the staggered magnetization $m$. Concurrently, the 
uniform orbital polarization saturates, $\tau^z=1$. We thus find that the 
magnetic transition appear simultaneously with the emptying of the narrow 
band, as expected by the Stoner instability of a half-filled single band.

\begin{figure}
 \centering \includegraphics[width=0.5\textwidth]{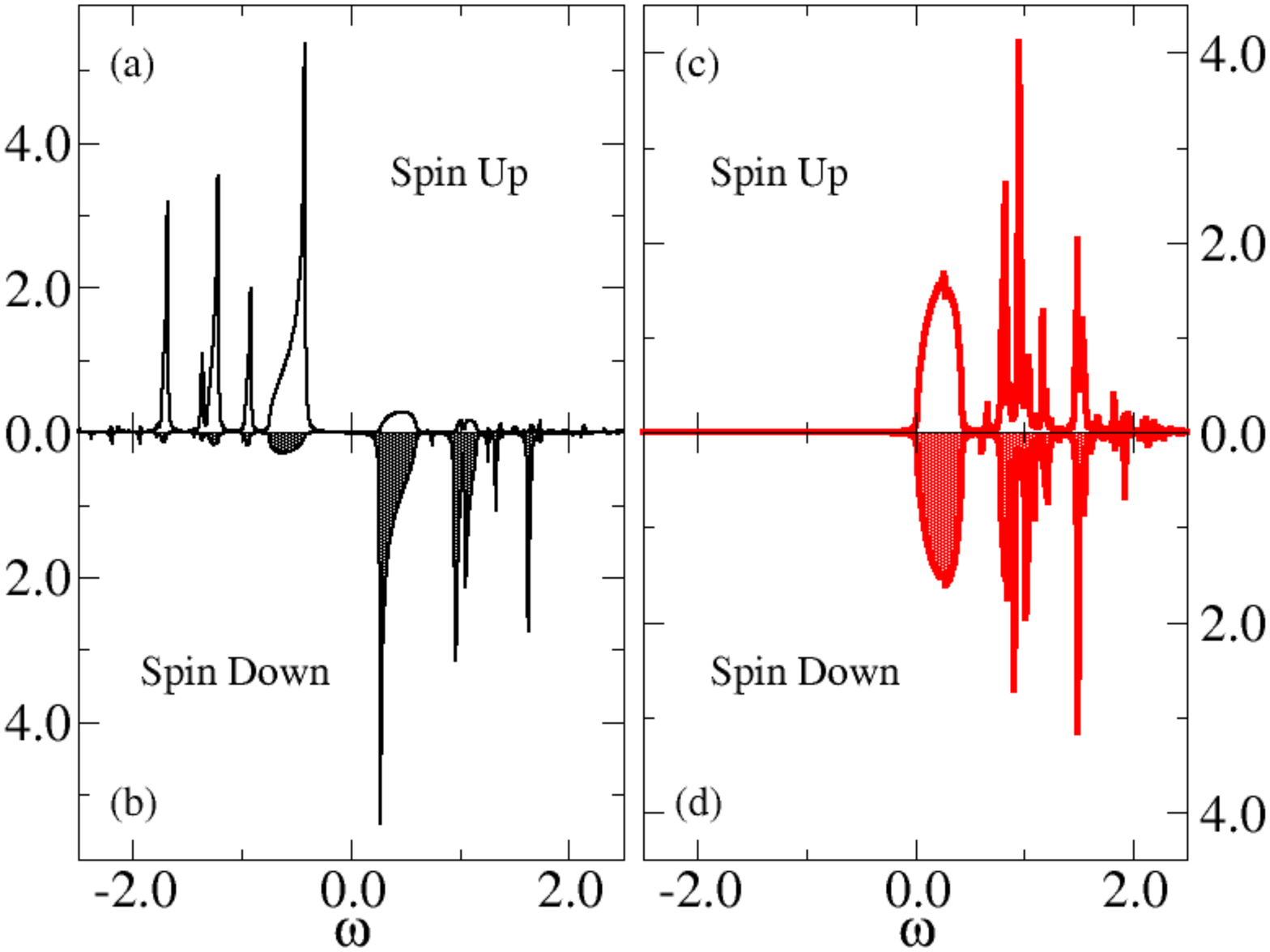}
  \caption{ (Color online) Spin resolved spectral
    functions for $\alpha = 0.4$, sub-lattice
    $\Lambda=A$, corresponding to majority spin up,  and $U = 1.6$. 
    Data for the wide band are in panels (a)-(b), those for the narrow
    band in panels (c)-(d). 
}
\label{fig_dos_spin_04_AFM}
\end{figure}

We can gain insight into the nature of the AFM phase at large
 $U$ by looking at the spin resolved spectral functions of the two orbitals, shown in \figu{fig_dos_spin_04_AFM}.
The wider band 1 has a particle-hole symmetric
spectrum. Conversely, the narrower band lies entirely above the Fermi level. 

\begin{figure}
 \centering \includegraphics[width=0.5\textwidth]{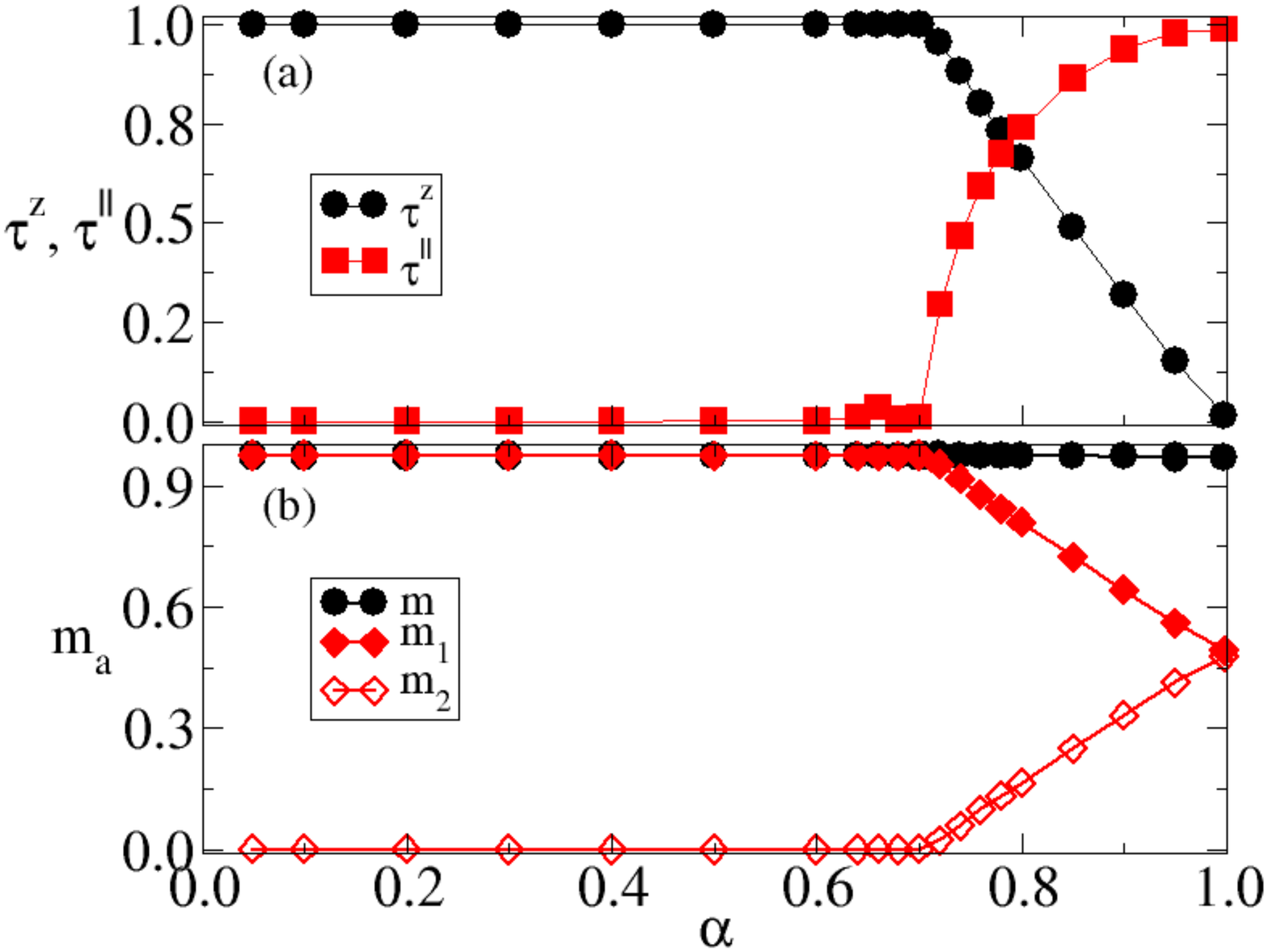}
  \caption{(Color online) (a) Uniform orbital polarization, $\tau^{z}$, and
    staggered one, $\tau^{||}$, as function of $\alpha$. 
    (b) Total and orbital resolved staggered magnetization,
    $m$, $m_{1}$ and $m_{2}$, as function of $\alpha$. 
    Data are for $U= 4.5$. 
    The solution displays a continuous  transition from the ferro-orbital antiferromagnetic state 
    to a canted antiferro-orbital but still antiferromagnetic state at $\alpha \simeq 0.7$. 
}
\label{fig_tau_magn_45_AFM} 
\end{figure}

We now study how the phase diagram changes with $\alpha$. 
In \figu{fig_tau_magn_45_AFM} we show the dependence upon $\alpha$ of 
the staggered magnetisation and polarisation, $m$ and $\tau^{||}$, 
respectively, and of the uniform orbital polarisation $\tau^z$, 
deep in the insulating phase at $U=4.5$. 
For $\alpha\lesssim 0.7$ we find the same behaviour as at $\alpha=0.4$, 
$m\simeq 1$, $\tau^z\simeq 1$ and $\tau^{||}=0$. 
Surprisingly, at $\alpha \simeq 0.7$ we observe a second order transition, 
above which also the orbital $U(1)$ symmetry breaks spontaneously and 
the model develops a finite staggered polarisation $\tau^{||}$. The staggered 
magnetisation remains almost saturated, but now has contribution from both 
bands. Indeed, since for $\alpha<1$ the solution corresponds to a canted AFO
ordering, the system has a finite FO component along the
$z$-direction of $\tau$, ultimately giving rise to AFM correlations similar
to the one-band case.

\begin{figure}
  \centering \includegraphics[width=0.5\textwidth]{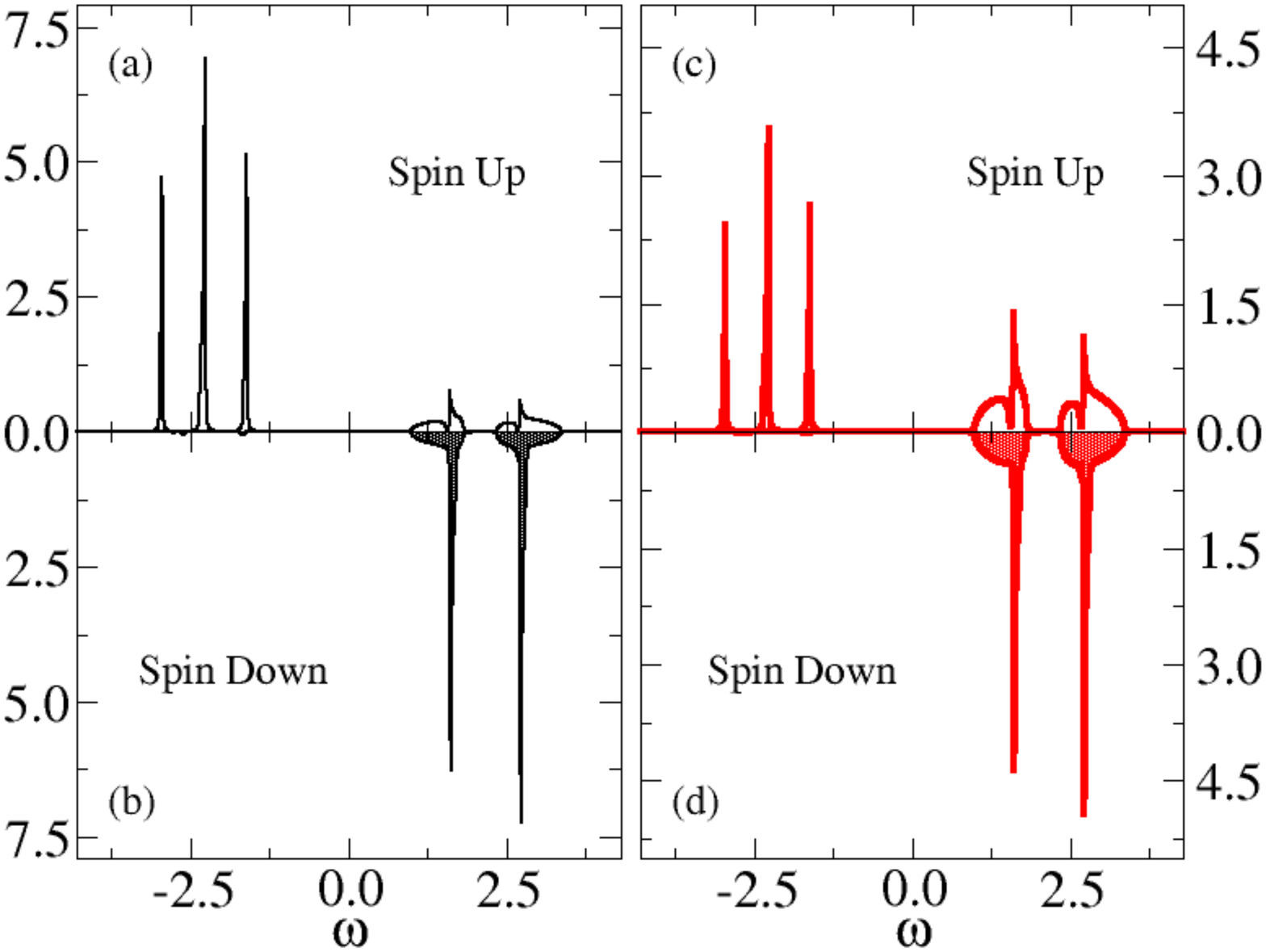}
  \caption{(Color online) Spin-resolved spectral functions for $\alpha = 0.9$ on
    sublattice $A$,  $U = 4.5$ for the wide band ((a)-(b)) and the
    narrow one ((c)-(d). 
 }
  \label{fig_dos_spin_45_AFM} 
\end{figure}
To get further insight in the nature of the AFM phase for $\alpha>0.7$ 
we show in  \figu{fig_dos_spin_45_AFM} the spin- and orbital-resolved spectral
functions at $\alpha=0.9$. It is instructive to compare these data with those 
reported in \figu{fig_dos_spin_04_AFM}. For this  larger value of the
bandwidth ratio, the two orbitals have almost indistinguishable spectral functions, unlike 
below the transition at $\alpha\simeq 0.7$.

\begin{figure}[bht]
 \centering \includegraphics[width=0.5\textwidth]{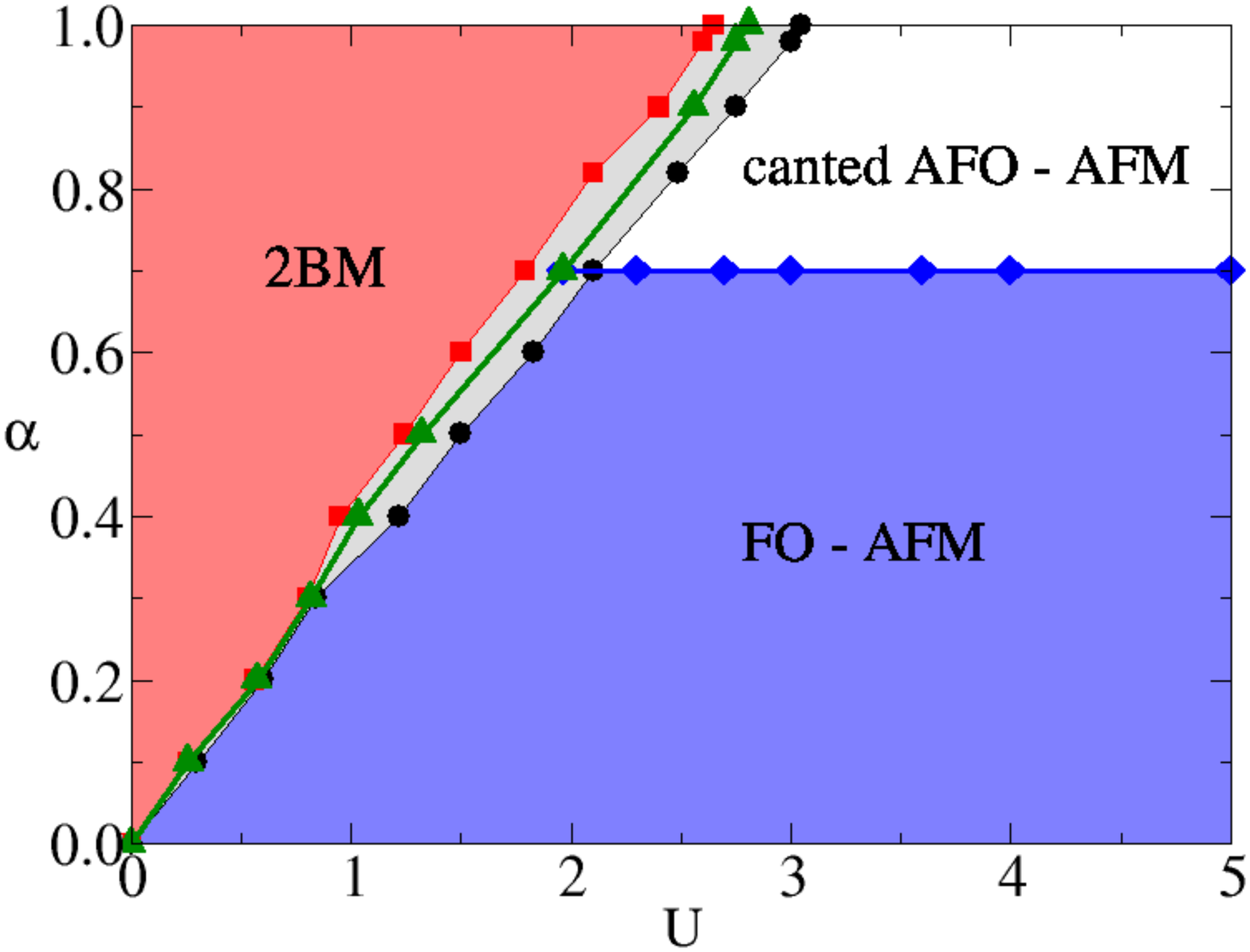}
  \caption{(Color online) Magnetic phase-diagram of the model in the
    $U$-$\alpha$ plane. 
    The phase diagram shows two main regions: a paramagnetic 2BM for
    small values of the interaction $U$ and an AFM insulator for
    $U>U^{AFM}_c$. The magnetic transition is of the first-order. The
    (gray) shaded area indicates the coexistence region. 
    The AFM phase is further divided in two by a continuous transition: an AFM with a canted AFO order for $\alpha>0.7$,  and an AFM with full orbital polarisation for $\alpha<0.7$. 
}
\label{fig_AFM} 
\end{figure}

We summarise our findings in the magnetic phase-diagram drawn in
\figu{fig_AFM}. We find three distinct phases. At small $U$ the 2BM is stable. 
For larger $U$ an AFM ordered insulator sets in. The magnetic transition is 
first-order, with a coexistence region that shrinks on approaching $\alpha=0$. 
The magnetic transition takes place for any $\alpha$ and for values of $U$ 
smaller than those required in the absence of magnetism, i.e. $U^{AFM}_c < U_c$. 
In particular, as expected by comparison with the single-band Hubbard
model, the 1BM region gets completely suppressed by the onset
of AFM order.  
Moreover, the AFM phase is cut in two by a second order 
transition line associated with a change in orbital ordering. 
For $\alpha<0.7$ the AFM has a saturated uniform orbital polarisation, in which 
only the wide band is occupied and contributes to the magnetic ordering. Increasing 
the bandwidth ratio above $\alpha\simeq 0.7$ leads to spontaneous orbital-$U(1)$ 
symmetry breaking, signalled by a finite in-plane staggered orbital polarisation. In 
this phase both bands are almost equally occupied and thus both contribute to the 
AFM order. Interestingly, we find that this transition is independent by the interaction 
strength $U$ and that we can reproduce it at the mean field level by assuming a 
value $\varepsilon \approx 0.7$ for the spin-spin correlation parameter that appears 
in \figu{fig_phase_diagr_strong_coupl}. 

We emphasise that the above results are valid as long as $\alpha<1$. When $\alpha=1$ the enlarged 
$SU(4)$ symmetry of the model may entail different type of spin-orbital orders\cite{PhysRevLett.107.215301} that we did not analyse.

\section{Conclusions}
\label{secV}

Despite its simplicity, two bands with different bandwidths subject to a 
monopole Slater integral $U$ and at quarter-filling, the model 
\eqn{hamiltonian} shows a remarkably rich phase diagram once the 
interplay between orbital and spin degrees of freedom are fully taken 
into account. In particular, because of the bandwidth difference, the 
interaction $U$ generates an effective crystal field that tends to empty 
the narrower band. This shows that correlations may not just enhance 
an existing crystal field, as pointed out in Ref.~\onlinecite{Poteryaev-PRB2007} 
in connection with the physics of V$_2$O$_3$, but even generate one 
despite its absence in the original Hamiltonian.  The depletion of the 
narrower band continues till a topological Lifshitz transition occurs, 
above which only the wider band remains occupied, and specifically 
half-filled. In our case study, with a bipartite lattice and unfrustrated 
Hamiltonian, as soon as the narrower band empties, a Stoner 
instability takes place driving the half-filled wider band into an 
antiferromagnetic insulator. This magnetic insulator still shows an 
active role of the orbital degrees of freedom that can drive a further 
phase transition between an insulator where only the wider band is 
occupied into another one where a canted antiferro-orbital order 
appears, and thus both bands are populated. The physics of the 
magnetic insulator observed at $\alpha \approx 1$ can describe 
some of the properties of the  KCuF$_3$ compound\cite{PhysRevLett.88.106403,PhysRevLett.101.266405}. 
However, we would like to emphasize that in the present work does not
take into account the strong  directionality of the $e_g$ bands in
$d$-orbitals compounds, which makes the comparison with realistic
materials hard. Such important effect is left for future work in this
direction. 

We argue that, in a  generic situation where some degree of frustration is unavoidably 
present, either geometric or caused by longer range hopping 
integrals, the one-band metal, with only the wider band occupied, 
might remain stable till a finite $U$ Mott transition, as we indeed 
found by preventing magnetism. We thus expect that the generic 
phase diagram must include, for not too strong repulsion $U$, a 
quarter-filled two-band metal separated by an interaction-induced 
Lifshitz transition from a half-filled one-band metal. Both metal 
phases must eventually give way to a Mott insulator above a critical 
$U$, whose precise magnetic and orbital properties will critically 
depend on the degree of frustration. We end emphasising that, at 
odds with the na{\"\i}ve expectation that a narrower band must also 
be the more correlated one, we here find right the opposite. This 
effect is due to the effective crystal field $\Delta^\mathrm{eff}$ that 
progressively empties the narrow band and at the same time brings 
the broad band closer and closer to the half filling condition,
enhancing the correlation effect on the wider band.

\section*{Acknowledgements}
We acknowledge support from the H2020 Framework Programme, under 
ERC Advanced Grant No. 692670 ``FIRSTORM''. 
A.A. and M.C. also acknowledge financial support from MIUR PRIN 2015
(Prot. 2015C5SEJJ001) and SISSA/CNR project "Superconductivity,
Ferroelectricity and Magnetism in bad metals" (Prot. 232/2015).

\bibliography{biblio}{}
\end{document}